\documentclass[journal]{IEEEtran}
\usepackage[colorlinks,linkcolor=magenta,anchorcolor=magenta,citecolor=magenta,bookmarks=true]{hyperref}  
\usepackage[T1]{fontenc}
\ifCLASSINFOpdf
\usepackage[pdftex]{graphicx} 
\DeclareGraphicsExtensions{.eps,.pdf,.jpeg,.png}
\else
\usepackage[dvips]{graphicx}
\fi
\usepackage[compress,nospace]{cite}
\usepackage[cmex10]{amsmath}
\usepackage{epstopdf,amsthm,stfloats,siunitx,amssymb,wasysym,algorithm,algorithmic,array,url, color,subfigure} 
\usepackage{footnote}
\makesavenoteenv{tabular}
\usepackage{soul}
\soulregister\cite7
\soulregister\ref7  

\begin{document}
 
\title{A RIS-Based Vehicle DOA Estimation Method With Integrated Sensing and Communication System}

\author{Zhimin~Chen,~\IEEEmembership{Member,~IEEE,} Peng~Chen,~\IEEEmembership{Senior~Member,~IEEE,} 
Ziyu~Guo,~\IEEEmembership{Member,~IEEE,} Yudong~Zhang,~\IEEEmembership{Senior~Member,~IEEE,} Xianbin~Wang,~\IEEEmembership{Fellow,~IEEE}
\thanks{This work was supported in part by the Natural Science Foundation of Shanghai under Grant 22ZR1425200, the Natural Science Foundation for Excellent Young Scholars of Jiangsu Province under Grant BK20220128, the National Key Laboratory of Wireless Communications Foundation under Grant IFN20230105, the National Key Laboratory on Electromagnetic Environmental Effects and Electro-optical Engineering under Grant JCKYS2023LD6, the Open Fund of ISN State Key Lab under Grant ISN24-04, the National Natural Science Foundation of China under Grant 61801112,	and the Shanghai Science and Technology Plan Project under Grant 21010501000. \textit{(Corresponding author: Peng Chen)}}
\thanks{Z.~Chen is with the School of Electronic and Information, Shanghai Dianji University, Shanghai 201306, China, and with State Key Laboratory of Integrated Services Networks, Xidian University, Xi'an 710071, China (e-mail: chenzm@sdju.edu.cn).}
\thanks{P.~Chen is with the State Key Laboratory of Millimeter Waves, Southeast University, Nanjing 210096, China, and also with State Key Laboratory of Integrated Services Networks, Xidian University, Xi'an 710071, China (e-mail: chenpengseu@seu.edu.cn).}
\thanks{Z.~Guo is with the School of Information Science and Technology, and also with the State Key Laboratory of Integrated Chips and Systems, Fudan University, Shanghai 200433, China (e-mail: zguo@fudan.edu.cn). }
\thanks{Y.~Zhang is with with the School of Computing and Mathematical Sciences, University of Leicester, Leicester, LE1 7RH, UK, and with School of Computer Science and Engineering, Southeast University, Nanjing 210096, China (e-mail: yudongzhang@ieee.org).}  
 \thanks{X.~Wang is with the Department of Electrical and Computer Engineering, Western University, London, ON N6A 5B9, Canada (e-mail: xianbin.wang@uwo.ca).}  
}
 
\markboth{IEEE Transactions on Intelligent Transportation Systems}%
{Shell \MakeLowercase{\textit{et al.}}: Bare Demo of IEEEtran.cls for IEEE Journals}
 
\maketitle
 
\begin{abstract}
With the development of intelligent transportation, growing attention has been received to integrated sensing and communication (ISAC) systems. In this paper, we formulate a novel passive sensing technique to obtain information on the vehicle's direction of arrival (DOA) using reconfigurable intelligent surfaces (RIS). A novel estimation method is proposed in the scenario with a receiver using only one full-functional channel, where multiple measurements for the DOA estimation are achieved by controlling the reflection matrix (measurement matrix) in the RIS. Moreover, different from the existing estimation methods, we also consider the interference signals introduced by wireless communication in the ISAC system. Then, we propose a novel atomic norm-based method to remove the interference signals and reconstruct the sparse signal. Additionally,  a novel Hankel-based multiple signal classification (MUSIC) method is formulated to obtain the DOA information after the interference removal. To reduce the interference signals more efficiently and improve the performance of the sparse reconstruction, we optimize the measurement matrix to improve the signal-to-interference-plus-noise ratio (SINR). Finally, the theoretical Cram'{e}r-Rao lower bound (CRLB)  is derived for the ISAC system on the vehicle DOA estimation. Simulation results show that the proposed method can achieve better performance in the DOA estimation, and the corresponding CRLB with different distributions of the sensing nodes are shown. The code for the proposed method is available online \url{https://github.com/chenpengseu/PassiveDOA-ISAC-RIS.git}.
\end{abstract}
 

\section{Introduction}
\IEEEPARstart{T}{he} integrated sensing and communication (ISAC)~\cite{9591331} system has attracted growing attention in recent years due to its potential  to meet the diverse needs of vehicle location-based applications~\cite{9843909,9963569,9500152}, and can provide both the sensing and wireless communication functions in the intelligent transportation. The principle of ISAC is similar to the joint communication and radar sensing (JCR) system~\cite{9540344}, where both the target sensing and wireless communication functions are concurrently realized~\cite{9585321,9668964,9667503,9634053}. In intelligent transportation using the ISAC system, the vehicles can be sensed either in an active way with a dedicated signal transmitted for sensing~\cite{9724252}, or in a passive way, where the estimation of the parameters, including the velocity and position, is achieved based on the reception of other signals unrelated to sensing (e.g., communication signals). In~\cite{9627227}, an overview of the modulation schemes for the JCR is given for the joint sensing and communication performance, and a comparative analysis is also carried out. Like the radar system, device-free sensing uses the reflected communication signal to sense the vehicle. The device-free sensing method for moving targets using orthogonal frequency division multiplexing (OFDM) is developed in~\cite{9724258}. When the intelligent transportation is combined with the ISAC, it can reduce the duplication of transmissions, devices, and infrastructure~\cite{du2023isacempowered}. In~\cite{9814586}, the radar sensing performance is analyzed in the 5G vehicle-to-everything (V2X) network. For the vehicular communications networks (VCN), the transportation scenarios are considered to use the ISAC system, and several transportation-specific cases are studied in~\cite{9830717}.

A sensor can be easily added to an existing wireless communication system for preliminary passive sensing. For example, a WiFi-based sensing method is proposed in~\cite{7875148} to estimate human activity nearby. In~\cite{9406387}, target localization with the one-bit analog-to-digital-converter (ADC) is proposed for the internet-of-things application. For the device-free localization, especially for the intelligent transportation application, authors of \cite{9210117} used the phase response shift to localize targets and improve the localization performance. Recently, reconfigurable intelligent surface (RIS) has been adopted in wireless communication for controlled reconfiguration of communication environment~\cite{8811733,jsan9010015,8247211}. With this development, RIS-based localization methods have also been exploited recently. In~\cite{9709801}, a near-field localization method is addressed using compressed sensing (CS), and authors of ~\cite{9593143} proposed a JCR system with RIS.

In the passive sensing system, the sensor receives conventional communication signals for vehicle localization based on the direction-finding or the direction of arrival (DOA) estimation~\cite{9616449,8516371}, which is also important in the global navigation satellite systems (GNSS) applications for the vehicle localization~\cite{7444122,9913344}. However, the resolution of traditional DOA estimation using the fast Fourier transformation (FFT) is limited. To improve the DOA estimation accuracy while keeping complexity low, super-resolution methods based on subspace have been proposed, such as the multiple signal classification (MUSIC) algorithm~\cite{music}, and estimation of signal parameters via rotational invariance techniques (ESPRIT)~\cite{32276}. 
In~\cite{9173575}, a deep network architecture SBLNet is proposed for the DOA estimation to achieve the super-resolution. With the development of the CS, sparse reconstruction methods have also been proposed for the DOA estimation~\cite{9585542,8537983,9521821}. In most CS-based methods, the spatial domain is discretized into grids, introducing the \emph{off-grid} error since the targets cannot be at the grids exactly~\cite{9511099,9326400}. To avoid the off-grid problem, the continuous domain methods are proposed for the sparse reconstruction. For example, an atomic norm-based minimization method is proposed for the DOA estimation in~\cite{9110826,9146196,7314978}. In~\cite{8537911}, a single snapshot super-resolution method is proposed for the DOA estimation in arbitrary array geometries. A fractional Fourier transform-based method is proposed in~\cite{9108285} by combining the ANM. Ref.~\cite{9384289} shows an ANM-based method using irregular Vandermonde decomposition. 

In this paper, we formulate a passive vehicle sensing system based on the DOA estimation using the RIS, which is controlled by a measurement matrix to realize the multiple measurements. A sensor with only one full-functional receiving channel estimates the DOA of targets. Unlike the existing sensing methods, we consider a passive sensing method, where the receiving signals are interfered with unwanted signals. Then, we proposed a novel atomic norm-based  method to remove the interference, and the DOA is estimated by a novel Hankel-based MUSIC method, where the spatial spectrum is estimated by fewer RIS measurements. Moreover,  a novel optimizing method for the measurement matrix is formulated to control the RIS by removing the interference signal and breaking the error platform of the DOA estimation. A random vector is introduced to ensure the randomness of RIS measurement. Finally, a theoretical Cram\'{e}r–Rao lower bound (CRLB) for the proposed passive sensing system is derived, and can be used to optimize the system, such as the locations of the wireless access point (AP), the RIS, and the sensor. 

The remainder of this paper is organized as follows. The RIS system model for vehicle DOA estimation is given in Section~\ref{system}. The proposed DOA estimation method is proposed in Section~\ref{method}. Then, the measurement matrix is optimized by the proposed in Section~\ref{optimize}. Simulation results are carried out in Section~\ref{simulation}, and Section~\ref{conclusion} concludes the paper. 

\textit{Notations:} Upper-case boldface letters denote matrices and lower-case boldface letters denote vectors. The  matrix Hermitian and transpose are denoted as $(\cdot)^\text{H}$ and $(\cdot)^\text{T}$, respectively.  $\mathcal{R}\{\cdot\}$ denotes the real part of a complex value, respectively. $\text{Tr}\{\cdot\}$ is the trace of a matrix.  $\|\cdot\|_1$ and $\|\cdot\|_2$ are  the $\ell_1$ and $\ell_2$ norms, respectively. $\mathcal{E}(\cdot)$ is the expectation operation. $j$ is defined as $j\triangleq\sqrt{-1}$. $\mathbb{C}$ is the set of complex numbers. 

\section{The RIS System Model for Vehicle DOA Estimation}\label{system}
\begin{figure}
	\centering
	\includegraphics[width=3.2in]{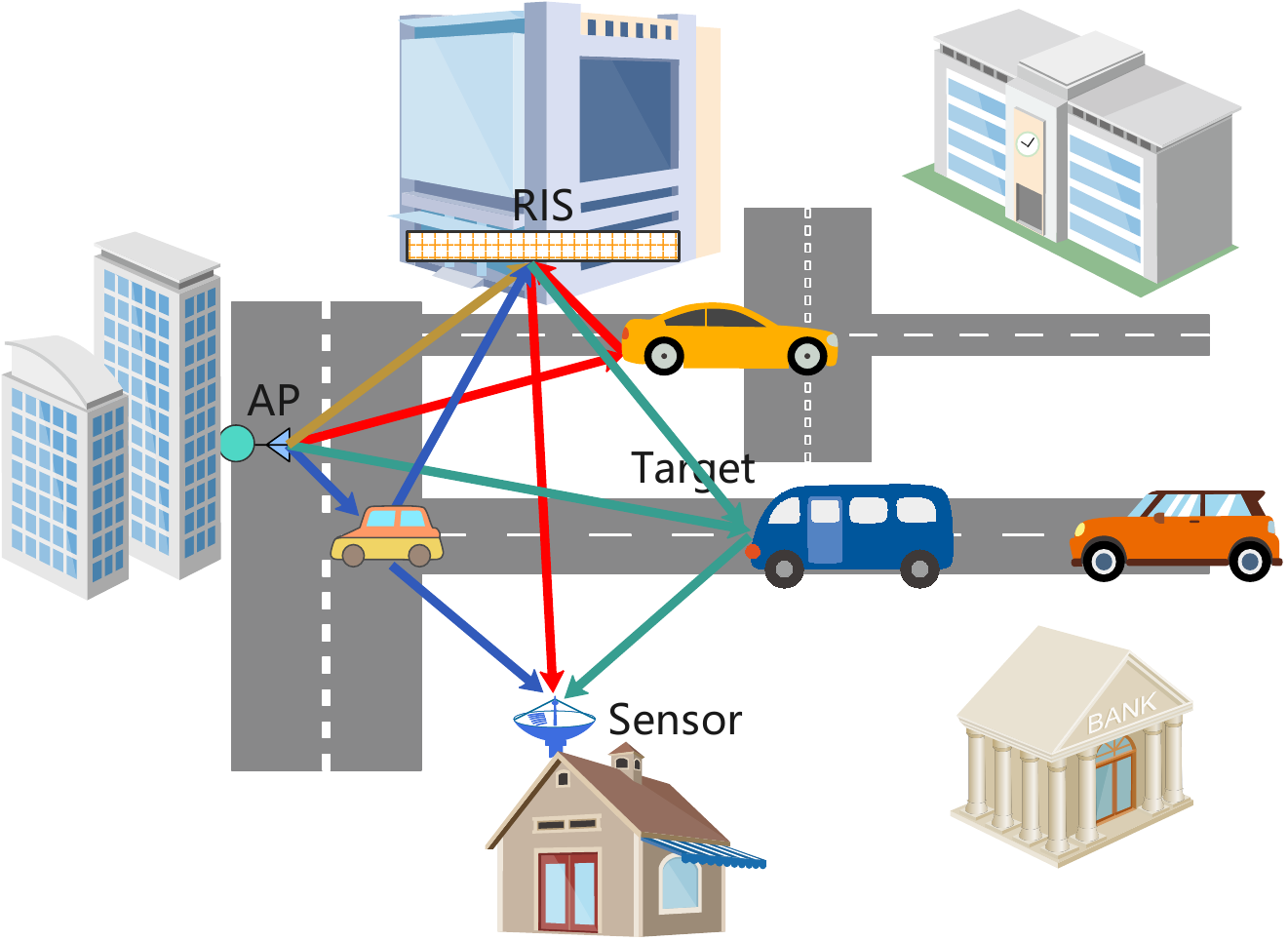}
	\caption{The vehicle sensing model using RIS and sensor with one full-functional receiver.}
	\label{sys}
\end{figure} 

This paper considers a  RIS-based sensing system to localize the vehicles in a scenario using wireless communication signals. As shown in Fig.~\ref{sys}, a wireless AP transmits wireless communication signals to the users. We use this signal to sense the vehicles using a receiving sensor with only one full-functional receiving channel. The full-functional receiving channel includes the filters, low noise amplifier, mixer, ADC and baseband processor, and has the ability to process the received signals. The RIS is equipped on the wall and can reflect all the received signals, and the transmitted signal in AP is denoted as $s(t)$. This paper considers a two-dimensional (2D) localization and can be easily extended to a three-dimensional (3D) one. Assume that there are $K$ vehicles, and the position of the $k$-th ($k=0,1,\dots, K-1$) vehicle is denoted as $\boldsymbol{p}_k=(x_k, y_k)$. The distance between the AP and the $k$-th vehicle is denoted as $d_{\text{AT},k}$, and the distance between the $k$-th vehicle and RIS is $d_{\text{TR},k}$. 

We consider a RIS with $M$ elements and assume that the elements form a uniform linear array (ULA) with the distance between adjacent elements being $d_{\text{E}}$. The DOA between the $k$-th vehicle and the RIS is denoted as $\theta_{\text{TR},k}$, and the DOA between the AP and the RIS is $\theta_{\text{AR}}$. Therefore, the superimposed signal received by the $m$-th ($m=0,1,\dots, M-1$) element in RIS can be expressed as
\begin{equation}\label{eq1}
\begin{split}
x_m(t) &= \sum^{K-1}_{k=0}\frac{\alpha_k}{d_{\text{AT},k}d_{\text{TR},k}}s(t-\tau_{\text{ATR},k})e^{j\frac{2\pi m d_{\text{E}}}{\lambda}\sin\theta_{\text{TR},k}}\\
&\quad +\frac{\beta}{d_{\text{AR}}}s(t-\tau_{\text{AR}})e^{j\frac{2\pi m d_{\text{E}}}{\lambda}\sin\theta_{\text{AR}}},   
\end{split} 
\end{equation}
where $\lambda$ is the wavelength, $\alpha_k$ is a constant corresponding to the scattering coefficient of the $k$-th vehicle, and $\beta$ is that of the direct path between the AP and RIS. $\tau_{\text{ATR},k}$ denotes the path delay from the AP to the $k$-th vehicle and received by the RIS. We denote the delay from the AP to the RIS as $\tau_{\text{AR}}$.  In (\ref{eq1}), the received signal $x_{m}(t)$ is a superimposed signal from $K$ vehicles, and $e^{j\frac{2\pi m d_{\text{E}}}{\lambda}\sin\theta_{\text{TR},k}}$ denotes the phase of the signal with the direction being $\theta_{\text{TR}, k}$ in the $m$-th element of the RIS. $e^{j\frac{2\pi m d_{\text{E}}}{\lambda}\sin\theta_{\text{AR}}}$ is the phase caused by the signal with the direction being $\theta_{\text{AR}}$.

During the the $n$-th ($n=0,1,\dots, N-1$) time slot $nT$, the reflected signal in the $m$-th ($m=0,1,\dots, M-1$) element is 
\begin{align}
    y_{m,n}(t) = A_{m,n}e^{j\phi_{m,n}}x_m(t),
\end{align}
where $A_{m,n}$ is the reflection amplitude and $\phi_{m,n}$ is the corresponding phase. Considering the direction between the RIS and sensor is $\theta_{\text{RS}}$, the received signal in the sensor with only one full-functional receiving channel can be obtained as 
\begin{align}
 	r_{n}(t)&=\sum^{M-1}_{m=0}\frac{\gamma}{d_{\text{RS}}}y_{m,n}(t-\tau_{\text{RS}})e^{j\frac{2\pi m d_{\text{E}}}{\lambda}\sin\theta_{\text{RS}}}+w_n(t)\notag\\
	&\qquad +\sum^{K-1}_{k=0}\frac{\alpha'_k}{d_{\text{AT},k}d_{\text{TS},k}}s(t-\tau_{\text{ATS},k})+\frac{\beta'}{d_{\text{AS}}}s(t-\tau_{\text{AS}})\notag\\
	&\qquad +\sum^{K-1}_{k=0}\sum^{M-1}_{m=0}\frac{\alpha''_k}{d_{\text{RT},k}d_{\text{TS},k}} y_{m,n}(t-\tau_{\text{RT},k}-\tau_{\text{TS},k})\notag\\
 &\qquad e^{j\frac{2\pi m d_{\text{E}}}{\lambda}\sin\theta_{\text{RT},k}},   
\end{align} 
where the delay between the RIS and the sensor is $\tau_{\text{RS}}$, and $w_{n}(t)$ is the additive white Gaussian noise (AWGN). The delay $\tau_{\text{ATS}}$ is that from the AP to the sensor through the vehicle, and the delay $\tau_{\text{AS}}$ is the direct path from the AP to the sensor. $d_{\text{RS}}$ is the distance between the RIS and the sensor. $\alpha'_k$, $\beta'$ and $\gamma$ are the  constants for the path and scattering attenuation. Note that the signals reflected by more times have been ignored since the power is much lower than that of other signals. Additionally, we assume that the size of RIS is much smaller than the bandwidth $B$ of the signal, so the delay between the elements in RIS can be described by the phase factor $e^{j\frac{2\pi m d_{\text{E}}}{\lambda}\sin\theta_{\text{RT},k}}$.

Finally, by ignoring the signal with much lower power, the received signal $r_n(t)$ can be simplified as
\begin{align}
r_n(t) &\approx  \sum^{M-1}_{m=0} \sum^{K-1}_{k=0} 
 A_{m,n}e^{j\phi_{m,n}}e^{j\frac{2\pi m d_{\text{E}}}{\lambda}\sin(\theta_{\text{RS}}+\theta_{\text{TR},k})}\notag\\
 &\qquad \frac{\gamma\alpha_k}{d_{\text{AT},k}d_{\text{TR},k}d_{\text{RS}}}s(t-\tau_{\text{ATRS},k}) \notag\\
&\qquad +\sum^{M-1}_{m=0} A_{m,n}e^{j\phi_{m,n}}e^{j\frac{2\pi m d_{\text{E}}}{\lambda}\sin(\theta_{\text{RS}}+\theta_{\text{AR}})}\notag\\
 &\qquad\frac{\gamma\beta}{d_{\text{AR}}d_{\text{RS}}}s(t-\tau_{\text{ARS}})+w_n(t)\notag\\
	&\qquad +\sum^{K-1}_{k=0}\frac{\alpha'_k}{d_{\text{AT},k}d_{\text{TS},k}}s(t-\tau_{\text{ATS},k})+\frac{\beta'}{d_{\text{AS}}}s(t-\tau_{\text{AS}})\notag\\
	&\qquad +\sum^{K-1}_{k=0}\sum^{M-1}_{m=0}\frac{\alpha''_k\beta }{d_{\text{AR}}d_{\text{RT},k}d_{\text{TS},k}} A_{m,n}e^{j\phi_{m,n}}\notag\\
 &\qquad s(t-\tau_{\text{ARTS},k})e^{j\frac{2\pi m d_{\text{E}}}{\lambda}(\sin\theta_{\text{RT},k}+\sin\theta_{\text{AR}})}
\end{align}
where we denote the delay $\tau_{\text{ATRS},k}\triangleq \tau_{\text{ATR},k}+\tau_{\text{RS}}$, $\tau_{\text{ARS}}\triangleq \tau_{\text{AR}}+\tau_{\text{RS}}$ and $\tau_{\text{ARTS},k}\triangleq \tau_{\text{AR}}+\tau_{\text{RT},k}+\tau_{\text{TS},k}$.

The expression of the received signal $r_n(t)$ shows that the received signal is composed of various signals with different power. To localize the vehicles and improve the power of interesting signals, we must design the receiving beam of the sensor so that the power in the path from the RIS to the sensor can be improved and other paths received by the sensor can be ignored. Hence, we have $0\approx a'_k\approx a''_k \approx \beta' \ll \gamma$. Then, the received signal can be simplified as
\begin{align}
r_n(t) &\approx \sum^{M-1}_{m=0} \sum^{K-1}_{k=0} 
 A_{m,n}e^{j\phi_{m,n}}e^{j\frac{2\pi m d_{\text{E}}}{\lambda}\sin(\theta_{\text{RS}}+\theta_{\text{TR},k})}\notag\\
 &\qquad \frac{\gamma\alpha_k}{d_{\text{AT},k}d_{\text{TR},k}d_{\text{RS}}}s(t-\tau_{\text{ATRS},k}) \notag\\
&\qquad +\sum^{M-1}_{m=0} A_{m,n}e^{j\phi_{m,n}}e^{j\frac{2\pi m d_{\text{E}}}{\lambda}\sin(\theta_{\text{RS}}+\theta_{\text{AR}})}\notag\\
 &\qquad\frac{\gamma\beta}{d_{\text{AR}}d_{\text{RS}}}s(t-\tau_{\text{ARS}})+w_n(t),
\end{align}
where the paths AP $\rightarrow$ vehicles $\rightarrow$ RIS $\rightarrow$ sensor and AP $\rightarrow$ RIS $\rightarrow$ sensor are kept.  
    
In the RIS-based sensing system, we assume that we can change the reflection coefficient $A_{m,n}e^{j\phi_{m,n}}$ with a frequency $1/T$, which is much faster than the signal bandwidth $B$, i.e., $BT\ll 1$. Therefore, collect the received signal into a vector with $N$ time slots, and we have
\begin{align}
	\boldsymbol{r}(t) & \triangleq \begin{bmatrix}
		r_0(t), r_1(t),\dots, r_{N-1}(t)
	\end{bmatrix}^{\text{T}}\\
& = \boldsymbol{w}(t)+\sum^{K-1}_{k=0}  \frac{\gamma\alpha_k}{d_{\text{AT},k}d_{\text{TR},k}d_{\text{RS}}}s(t-\tau_{\text{ATRS},k})\notag\\
 &\qquad
\sum^{M-1}_{m=0} e^{j\frac{2\pi m d_{\text{E}}}{\lambda}\sin(\theta_{\text{RS}}+\sin\theta_{\text{TR},k})} \boldsymbol{g}_m\notag\\
&\qquad  + \frac{\gamma\beta}{d_{\text{AR}}d_{\text{RS}}}s(t-\tau_{\text{ARS}}) \sum^{M-1}_{m=0} e^{j\frac{2\pi m d_{\text{E}}}{\lambda}\sin(\theta_{\text{RS}}+\theta_{\text{AR}})}\boldsymbol{g}_m\notag\\
& = \sum^{K-1}_{k=0}  \frac{\gamma\alpha_k}{d_{\text{AT},k}d_{\text{TR},k}d_{\text{RS}}}s(t-\tau_{\text{ATRS},k})\boldsymbol{G}\boldsymbol{a}(\theta_{\text{TR},k})\notag\\
&\qquad  + \frac{\gamma\beta}{d_{\text{AR}}ds_{\text{RS}}}s(t-\tau_{\text{ARS}})
\boldsymbol{G}\boldsymbol{a}(\theta_{\text{AR}}) +\boldsymbol{w}(t)\notag
\end{align}
where we define $\boldsymbol{w}(t)\triangleq \begin{bmatrix}
	w_0(t),w_1(t),\dots, w_{N-1}(t)
\end{bmatrix}^{\text{T}}$, $g_{m,n}\triangleq A_{m,n}e^{j\phi_{m,n}}$, $\boldsymbol{g}_m\triangleq \begin{bmatrix}
	g_{m,0},g_{m,1},\dots, g_{m,N-1}
\end{bmatrix}^{\text{T}}$, and the measurement matrix is defined as $\boldsymbol{G}\triangleq \begin{bmatrix}
	\boldsymbol{g}_0,\boldsymbol{g}_1,\dots, \boldsymbol{g}_{M-1}
\end{bmatrix}\in\mathbb{C}^{N\times M}$. We also define the steering vector as
\begin{align}
\boldsymbol{a}(\theta)\triangleq\begin{bmatrix}
1, e^{j\frac{2\pi d_{\text{E}}}{\lambda}\sin(\theta_{\text{RS}}+\theta)}, \dots, e^{j\frac{2\pi (M-1) d_{\text{E}}}{\lambda}\sin(\theta_{\text{RS}}+\theta)}
\end{bmatrix}^{\text{T}}. 
\end{align}

By defining $z_k(t)\triangleq \frac{\gamma\alpha_k}{d_{\text{AT},k}d_{\text{TR},k}d_{\text{RS}}}s(t-\tau_{\text{ATRS},k})$, we have $\boldsymbol{z}(t)\triangleq \begin{bmatrix}
	z_0(t),z_1(t),\dots, z_{K-1}(t)
\end{bmatrix}^{\text{T}}$. Additionally, we also define $q(t)\triangleq \frac{\gamma\beta}{d_{\text{AR}}d_{\text{RS}}}s(t-\tau_{\text{ARS}})
$. Finally, the received signal in the sensor can be simplified in a vector form
\begin{align}
\boldsymbol{r}(t) = \boldsymbol{G}\boldsymbol{A}(\boldsymbol{\theta}_{\text{TR}})\boldsymbol{z}(t)	+\boldsymbol{G}\boldsymbol{a}(\theta_{\text{AR}})q(t)+\boldsymbol{w}(t),\label{eq8}
\end{align}
where we define
\begin{align}
	\boldsymbol{A}(\boldsymbol{\theta}_{\text{TR}})\triangleq \begin{bmatrix}
	\boldsymbol{a}(\theta_{\text{TR},0}), \boldsymbol{a}(\theta_{\text{TR},1}),\dots, \boldsymbol{a}(\theta_{\text{TR},K-1})
\end{bmatrix}\in\mathbb{C}^{M\times K}.
\end{align}
 
In the system model (\ref{eq8}), since the passive sensing method is considered, the measurement matrix $\boldsymbol{G}$ in the RIS and the received signal $\boldsymbol{r}(t)$ are known. However, the signals $\boldsymbol{z}(t)$ and the interference $q(t)$ are unknown. We are trying to estimate the DOA vector $\boldsymbol{\theta}_{\text{TR}}$ from the received signal $\boldsymbol{r}(t)$.

\section{The Sparse-Based Vehicle DOA Estimation Method}\label{method}
\subsection{The Atomic Norm Minimization-Based Denoising Method}
From the received signal (\ref{eq8}), we can find that the vehicle echoed signals are interfered with the direct signal from the AP to the RIS, and the direct signal is also unknown in the passive sensing system. Therefore, we propose a novel method based on the ANM with interference cancellation. The parameter $t$ has been ignored for simplification in the following contents. The sensing problem can be formulated as
\begin{align}
\min_{\boldsymbol{\xi},\eta}\, 
\|\boldsymbol{r}-\boldsymbol{G\xi} -\boldsymbol{Ga}(\theta_{\text{AR}})\eta\|^2_2+\rho\|\boldsymbol{\xi}\|_{\mathcal{A}},\label{eq35}
\end{align}
where $\eta\in\mathbb{C}$, $\boldsymbol{\xi}\in\mathbb{C}^{M\times 1}$, and $\|\cdot\|^2_2$ denotes the square of $\ell_2$ norm. In practical, the value of $\rho$ can be chosen as  $\rho=\sigma_{\text{w}}\sqrt{M\log M}$~\cite{zheng_adaptive_2018}.

The optimization problem (\ref{eq35}) can be rewritten as
\begin{equation}
\begin{split}
\min_{\boldsymbol{\xi},\eta}\, & \|\boldsymbol{\xi}\|_{\mathcal{A}}\\
\text{s.t.}\, & \|\boldsymbol{r}-\boldsymbol{G\xi} -\boldsymbol{Ga}(\theta_{\text{AR}})\eta\|^2_2 \leq \rho'.   
\end{split}    
\end{equation}
The corresponding Lagrangian function with dual variable $\bar{\rho}$ can be expressed as
\begin{align}
& \mathcal{L}(\boldsymbol{\xi}, \eta, \bar{\rho})  =   \|\boldsymbol{\xi}\|_{\mathcal{A}}+\bar{\rho} (\|\boldsymbol{r}-\boldsymbol{G\xi} -\boldsymbol{Ga}(\theta_{\text{AR}})\eta\|^2_2 - \rho')\notag\\
&\quad = \|\boldsymbol{r}-\boldsymbol{G\xi} -\boldsymbol{Ga}(\theta_{\text{AR}})\eta\|^2_2+1/\bar{\rho} (\|\boldsymbol{\xi}\|_{\mathcal{A}} - \bar{\rho}\rho')
\end{align}

To solve the optimization problem (\ref{eq35}), the ANM-based method is proposed, where $\|\boldsymbol{\xi}\|_{\mathcal{A}}$ is the atomic norm of $\boldsymbol{\xi}$, and is defined as
\begin{align}
\|\boldsymbol{\xi}\|_{\mathcal{A}}\triangleq \inf\left\{c\geq 0: \boldsymbol{\xi}\in c\cdot \operatorname{conv}\{\mathbb{A}\}\right\}.
\end{align}
The atomic norm denotes a smallest nonnegative scaling of $\operatorname{conv}\{\mathbb{A}\}$ until it intersects, and a fundamental geometric property is an atomic norm ball. $\mathbb{A}$ is an atomic set for the DOA estimation and is defined as
\begin{align}
\mathbb{A} = \left\{e^{j\phi_{\text{A}} } \boldsymbol{a}(\theta):\theta\in[-\pi/2, \pi/2],  \phi_{\text{A}}\in[0, 2\pi)\right\}.
\end{align}

The optimization problem cannot be solved directly. We first obtain the following atomic norm expression
\begin{align}
\|\boldsymbol{\xi}\|_{\mathcal{A}} = \inf \left\{\sum_p  c_p: \boldsymbol{\xi}=\sum_p c_p \boldsymbol{a}_p, c_p\geq 0, \boldsymbol{a}_p\in \mathbb{A} \right\}.
\end{align}

According to ~\cite{chi_harnessing_2020,7313018,6576276}, the atomic norm can also be obtained by an equivalent semidefinite program (SDP) problem
\begin{align}
\|\boldsymbol{\xi}\|_{\mathcal{A}}=\inf_{\boldsymbol{u},\nu}\bigg\{&
\frac{1}{2N}\operatorname{Tr}\{\operatorname{Toep}(\boldsymbol{u})\}+\frac{1}{2}\nu: \notag\\
&\quad \begin{bmatrix}
\operatorname{Toep}(\boldsymbol{u}), & \boldsymbol{\xi}\\
\boldsymbol{\xi}^{\text{H}}, & \nu 
\end{bmatrix}\succeq 0
\bigg\},
\end{align}
where $\boldsymbol{u}\in\mathbb{C}^{M\times 1}$ and $\operatorname{Toep}(\boldsymbol{u})$ denotes a Toeplitz matrix with the first column being $\boldsymbol{u}$ and can be expressed as
\begin{align}
\operatorname{Toep}(\boldsymbol{u})\triangleq\begin{bmatrix}
u_0 &u_{-1}&\dots& u_{-(M-1)}\\
u_1 &u_0 &\dots& u_{-(M-2)}\\
\vdots& \vdots&\ddots & \vdots\\
u_{M-1} &u_{M-2}& \dots & u_0
\end{bmatrix}.    
\end{align}
 
Hence, by defining $\boldsymbol{b}\triangleq \boldsymbol{Ga}(\theta_{\text{AR}})\in\mathbb{C}^{N \times 1}$, the optimization problem in (\ref{eq35}) can be rewritten as
\begin{align}
\min_{\boldsymbol{\xi},\boldsymbol{u},\boldsymbol{\nu},\eta}\, &
\|\boldsymbol{r}-\boldsymbol{G\xi} -\eta\boldsymbol{b}\|^2_2+\frac{\rho}{2}\left[\frac{1}{M}\operatorname{Tr}\{\operatorname{Toep}(\boldsymbol{u})\}+\nu\right]\notag\\
\text{s.t.}\,& \begin{bmatrix}
\operatorname{Toep}(\boldsymbol{u}), & \boldsymbol{\xi}\\
\boldsymbol{\xi}^{\text{H}}, & \nu 
\end{bmatrix}\succeq 0.\label{eq42}
\end{align}
The denoising vector $\boldsymbol{\xi}$ can be obtained by solving the SDP problem.

\subsection{The Hankel-Based MUSIC Method for the Spatial Spectrum Estimation}
With the SDP problem (\ref{eq42}), we can reconstruct the signal $\boldsymbol{\xi}$ with the sparsity priority. However, the signal $\boldsymbol{\xi}$ is sparse in the frequency domain, and the DOA vector $\boldsymbol{\theta_{\text{TR}}}$ is still unknown. We will use a MUSIC-based method to estimate the DOA without discretizing the spatial domain into grids. Since the reconstruction signal $\boldsymbol{\xi}\in\mathbb{C}^{M\times 1}$ is a vector, the covariance matrix cannot be obtained with only one snapshot. The traditional MUSIC cannot be used directly, so a Hankel matrix-based method is proposed to realize a MUSIC-based method~\cite{liao_music_2016}. First, a Hankel matrix is obtained as
\begin{align}
\text{Hankel}(\boldsymbol{\xi})=\begin{bmatrix}
\boldsymbol{\xi}^{\text{T}}_{0:M-L}\\
\boldsymbol{\xi}^{\text{T}}_{1:1+M-L}\\
\vdots\\
\boldsymbol{\xi}^{\text{T}}_{L-1:M-1}\\
\end{bmatrix},
\end{align}
where $\boldsymbol{\xi}_{m':n'}$ denotes a sub-vector of $\boldsymbol{\xi}$ using the $m'$-th to the $n'$-th entries. We reshape a vector $\boldsymbol{\xi}$ using $M$ elements to a matrix using $L$ elements.

Then, a singular value decomposition (SVD) is adopted to obtain the noise subspace, and we have
\begin{align}
\text{Hankel}(\boldsymbol{\xi})=\boldsymbol{U}_1\boldsymbol{D}\boldsymbol{U}_2,
\end{align}
where $\boldsymbol{D}$ is a matrix with the diagonal entries being the singular values. $\boldsymbol{U}_1$ and $\boldsymbol{U}_2$ are the complex unitary matrices. The columns of $\boldsymbol{U}_1$ and $\boldsymbol{U}_2$ are the left-singular and right-singular vectors, respectively. 

The noise subspace can be obtained as $\Tilde{\boldsymbol{U}}$ from the columns of $\boldsymbol{U}_1$ corresponding to the small singular values. Finally, the spatial spectrum with only one snapshot can be obtained as 
\begin{align}
g_{\text{sp}}(\theta) =\frac{\|\boldsymbol{\Tilde{a}}(\theta)\|^2_2
}{\|\boldsymbol{\Tilde{a}}^{\text{H}}(\theta)\Tilde{\boldsymbol{U}}\|^2_2},
\end{align}
where we define a steering vector for sub-array as
\begin{align}
\boldsymbol{\Tilde{a}}(\theta)\triangleq\begin{bmatrix} 
1, e^{j\frac{2\pi d_{\text{E}}}{\lambda}\sin(\theta_{\text{RS}}+\theta)}, \dots, e^{j\frac{2\pi (L-1) d_{\text{E}}}{\lambda}\sin(\theta_{\text{RS}}+\theta)}
\end{bmatrix}^{\text{T}}.  
\end{align}

The DOA $\boldsymbol{\theta_{\text{TR}}}$ can be estimated by finding the peak values of $g_{\text{sp}}(\theta)$. The details of the proposed method are summarized in Algorithm~\ref{alg1}. The computational complexity of the proposed method is determined by the SDP and SVD steps. The computational complexity of SDP step is about $\mathcal{O}((M+1)^4)$, and that of SVD is about $\mathcal{O}(L^2(M-L+1)+M(M-L+1)^2+(M-L+1)^3)$. Since $M>L$, the computational complexity of Algorithm~\ref{alg1} is about $\mathcal{O}((M+1)^4)$.

\begin{algorithm}
	\caption{ANM-PDOA: Atomic norm-based passive DOA estimation method for the RIS-based sensing system} \label{alg1}
	\begin{algorithmic}[1]
		\STATE  \emph{Input:} The received signal $\boldsymbol{r}$, the measurement matrix $\boldsymbol{G}$, the direction $\theta_{\text{RS}}$ between the RIS  and the sensor, the direction $\theta_{\text{AR}}$ between the AP  and the RIS, the number of RIS elements $M$, the wavelength $\lambda$, the distance between the adjacent RIS elements $d_{\text{E}}$, and the size $L$ of sub-array.
		\STATE Define the steering vector as
		\begin{align}
		    \boldsymbol{a}(\theta)\triangleq\begin{bmatrix}
                1, \dots, e^{j\frac{2\pi (M-1) d_{\text{E}}}{\lambda}\sin(\theta_{\text{RS}}+\theta)}
                \end{bmatrix}^{\text{T}}. 
		\end{align}
		\STATE $\boldsymbol{b}=\boldsymbol{Ga}(\theta_{\text{AR}})$.
		\STATE Obtain $\boldsymbol{\xi}$ from
    		\begin{align}
            \min_{\boldsymbol{\xi},\boldsymbol{u},\boldsymbol{\nu},\eta}\, &
            \|\boldsymbol{r}-\boldsymbol{G\xi} -\eta\boldsymbol{b}\|^2_2+\frac{\rho}{2}\left[\frac{1}{M}\operatorname{Tr}\{\operatorname{Toep}(\boldsymbol{u})\}+\nu\right]\notag\\
            \text{s.t.}\,& \begin{bmatrix}
            \operatorname{Toep}(\boldsymbol{u}), & \boldsymbol{\xi}\\
            \boldsymbol{\xi}^{\text{H}}, & \nu 
            \end{bmatrix}\succeq 0. 
            \end{align}
        \STATE Obtain a Hankel matrix from 
            \begin{align}
            \text{Hankel}(\boldsymbol{\xi})=\begin{bmatrix}
            \boldsymbol{\xi}_{0:M-L},
            \dots,
            \boldsymbol{\xi}_{L-1:M-1}
            \end{bmatrix}^{\text{T}}.
            \end{align}
        \STATE The SVD is used as $\text{Hankel}(\boldsymbol{\xi})=\boldsymbol{U}_1\boldsymbol{D}\boldsymbol{U}_2$.
        \STATE Obtain the noise subspace can be obtained as $\Tilde{\boldsymbol{U}}$ from the columns of $\boldsymbol{U}_1$ corresponding to the small singular values.
        \STATE Define the steering vector of a sub-array as 
            \begin{align}
            \boldsymbol{\Tilde{a}}(\theta)\triangleq\begin{bmatrix} 
            1, \dots, e^{j\frac{2\pi (L-1) d_{\text{E}}}{\lambda}\sin(\theta_{\text{RS}}+\theta)}
            \end{bmatrix}^{\text{T}}.  
            \end{align}
        \STATE The spatial spectrum is obtained as 
            \begin{align}
            g_{\text{sp}}(\theta) =\frac{\|\boldsymbol{\Tilde{a}}(\theta)\|^2_2
            }{\|\boldsymbol{\Tilde{a}}^{\text{H}}(\theta)\Tilde{\boldsymbol{U}}\|^2_2},
            \end{align}
		\STATE The DOA $\hat{\boldsymbol{\theta}}_{\text{TR}}$ can be estimated from the peak values of $g_{\text{sp}}(\theta)$.
		\STATE \emph{Output:} The estimated DOA $\hat{\boldsymbol{\theta}}_{\text{TR}}$.
	\end{algorithmic}
\end{algorithm}

\section{Optimizing the measurement matrix}\label{optimize}
From the system model (\ref{eq8}), we can find that the SINR of the received signal can be expressed as
\begin{align}
g_{\text{SINR}}(\boldsymbol{G}) = \frac{\mathcal{E}\left\{\|\boldsymbol{G}\boldsymbol{A}(\boldsymbol{\theta}_{\text{TR}})\boldsymbol{z}(t)\|^2_2\right\}}{\mathcal{E}\left\{\|\boldsymbol{G}\boldsymbol{a}(\theta_{\text{AR}})q(t)\|^2_2\right\}+\mathcal{E}\|\boldsymbol{w}(t)\|^2_2},
\end{align}
which can be simplified as 
\begin{align}
g_{\text{SINR}}(\boldsymbol{G}) = \frac{
\operatorname{Tr}\left\{\boldsymbol{G}\boldsymbol{A}(\boldsymbol{\theta}_{\text{TR}})
\boldsymbol{\Lambda}\boldsymbol{A}^{\text{H}}(\boldsymbol{\theta}_{\text{TR}})
\boldsymbol{G}^{\text{H}}
\right\}
}{\|\boldsymbol{G}\boldsymbol{a}(\theta_{\text{AR}})\|^2_2\left|\frac{\gamma\beta}{d_{\text{AR}}d_{\text{RS}}}\right|^2P_s+N\sigma^2_{\text{w}}},
\end{align}
and $\boldsymbol{\Lambda}\triangleq \gamma^2P_\text{s} \operatorname{diag}\left\{\left\|\frac{\alpha_0}{d_{\text{AT,0}}d_{\text{TR,0}}d_{\text{RS}}}\right|^2,\dots, \left\|\frac{\alpha_{K-1}}{d_{\text{AT,K-1}}d_{\text{TR,K-1}}d_{\text{RS}}}\right|^2\right\}$

Hence, we can optimize the measurement matrix $\boldsymbol{G}$ to improve the SINR of the received signal. In the practical sensing system, the vehicles can be in any direction, so the DOA vector $\boldsymbol{\theta_{\text{TR}}}$ is unknown. To improve the SINR of the received signal, we can decrease the power from the AP so that the  following optimization problem can be formulated 
\begin{align}
\min_{\boldsymbol{G}}\, & \|\boldsymbol{G}\boldsymbol{a}(\theta_{\text{AR}})\|^2_2\left|\frac{\gamma\beta}{d_{\text{AR}}d_{\text{RS}}}\right|^2P_s+N\sigma^2_{\text{w}}\label{eq12}\\
\text{s.t.}\, & |g_{m,n}|\leq 1.&\notag
\end{align}

To solve the optimization problem (\ref{eq12}), we rewrite the measurement matrix as $\boldsymbol{G}=[\boldsymbol{g}'_0,\boldsymbol{g}'_1,\dots,\boldsymbol{g}'_{N-1}]^{\text{H}}$, so we can optimize the row of $\boldsymbol{G}$ iterative. For the $n$-th row, we can formulate the following problem
\begin{align}
\min_{\boldsymbol{g}'_n}\, & \|\boldsymbol{g}'^{\text{H}}_n\boldsymbol{a}(\theta_{\text{AR}})\|^2_2\label{eq13}\\
\text{s.t.}\, & |g'_{m,n}|\leq 1.&\notag
\end{align}
Since we have 
\begin{align}
\|\boldsymbol{g}'^{\text{H}}_n\boldsymbol{a}(\theta_{\text{AR}})\|^2_2=\operatorname{Tr}\{\boldsymbol{a}(\theta_{\text{AR}})\boldsymbol{a}^{\text{H}}(\theta_{\text{AR}})\boldsymbol{g}'_n\boldsymbol{g}'^{\text{H}}_n\},    
\end{align}
the following SDP problem can approximate the optimization problem (\ref{eq13})
\begin{equation}
\begin{split}
\min_{\Tilde{\boldsymbol{G}}}\, & \operatorname{Tr}\{\boldsymbol{a}(\theta_{\text{AR}})\boldsymbol{a}^{\text{H}}(\theta_{\text{AR}})\Tilde{\boldsymbol{G}}\}\\
\text{s.t.}\, & \Tilde{G}_{m,m}=1\,(m=0,1,\dots,M-1) \\
& \Tilde{\boldsymbol{G}}\succeq 0\\
& \Tilde{\boldsymbol{G}}\text{ is a Hermitian matrix},     
\end{split}    \label{eq15}
\end{equation}
where $\Tilde{G}_{m,m}$ denotes the $m$-th diagonal entry of $\Tilde{\boldsymbol{G}}$, and we force on optimizing the phases of the RIS elements.

The problem (\ref{eq15}) is convex and can be solved efficiently, and we can obtain a matrix $\Tilde{\boldsymbol{G}}$. Since the measurement matrix $\boldsymbol{G}$ must have the random characteristic to guarantee the sparse reconstruction performance. With eigenvalue decomposition~\cite{9198125}, we have
\begin{align}
\Tilde{\boldsymbol{G}}=\Tilde{\boldsymbol{U}}\Tilde{\boldsymbol{\Lambda}}\Tilde{\boldsymbol{U}}^{\text{H}},
\end{align}
and the $n$-th row can be chosen as
\begin{align}
\boldsymbol{g}'_n = e^{j\operatorname{ang}\{\Tilde{\boldsymbol{U}}\Tilde{\boldsymbol{\Lambda}}^{\frac{1}{2}}\Tilde{\boldsymbol{g}}\}},
\end{align}
where $\text{ang}(\cdot)$ obtains the angle of a variable, and $\Tilde{\boldsymbol{g}}$ is a random vector with the entries following the complex Gaussian distribution. The details to optimize the measurement matrix are given in Algorithm~\ref{alg2}.

From the details of Algorithm~\ref{alg2}, we can find that the computational complexity is determined by solving the SDP problem. Hence, the computational complexity of the proposed method is about $\mathcal{O}(M^4)$, where $M$ is the number of the RIS elements.

\begin{algorithm}
	\caption{The optimization method for the measurement matrix} \label{alg2}
	\begin{algorithmic}[1]
		\STATE  \emph{Input:} The direction $\theta_{\text{RS}}$ between the RIS  and the sensor, the direction $\theta_{\text{AR}}$ between the AP  and the RIS, and the number of RIS elements $M$.
		\STATE Obtain the steering vector $\boldsymbol{a}(\theta_{\text{AR}})$ as
            \begin{align}
            \boldsymbol{a}(\theta_{\text{AR}})\triangleq\begin{bmatrix}
            1, \dots, e^{j\frac{2\pi (M-1) d_{\text{E}}}{\lambda}\sin(\theta_{\text{RS}}+\theta_{\text{AR}})}
            \end{bmatrix}^{\text{T}}. 
            \end{align} 
		\STATE Obtain a Hermitian matrix $\Tilde{\boldsymbol{G}}$ from 
            \begin{align}
            \min_{\Tilde{\boldsymbol{G}}}\, & \operatorname{Tr}\{\boldsymbol{a}(\theta_{\text{AR}})\boldsymbol{a}^{\text{H}}(\theta_{\text{AR}})\Tilde{\boldsymbol{G}}\}\\
            \text{s.t.}\, & \Tilde{G}_{m,m}=1\,(m=0,1,\dots,M-1)\notag \\
            & \Tilde{\boldsymbol{G}}\succeq 0\notag\\
            & \Tilde{\boldsymbol{G}}\text{ is a Hermitian matrix}. \notag
            \end{align}
        \STATE Use the eigenvalue decomposition $\Tilde{\boldsymbol{G}}=\Tilde{\boldsymbol{U}}\Tilde{\boldsymbol{\Lambda}}\Tilde{\boldsymbol{U}}^{\text{H}}$.  
        \STATE Generate a random vector $\Tilde{\boldsymbol{g}}$ following the complex Gaussian distribution.
        \STATE Obtain a vector $\boldsymbol{g}'_n$ ($n=0,1,\dots, N-1$) as
            \begin{align}
            \boldsymbol{g}'_n = e^{j\operatorname{ang}\{\Tilde{\boldsymbol{U}}\Tilde{\boldsymbol{\Lambda}}^{\frac{1}{2}}\Tilde{\boldsymbol{g}}\}}.
            \end{align}
        \STATE Generate the measurement matrix as
        \begin{align}
            \boldsymbol{G}=[\boldsymbol{g}'_0,\boldsymbol{g}'_1,\dots,\boldsymbol{g}'_{N-1}]^{\text{H}}.
        \end{align}
		\STATE \emph{Output:} The measurement matrix $\boldsymbol{G}$.
	\end{algorithmic}
\end{algorithm}

\section{The CRLB for the Passive DOA Estimation With RIS}\label{crlb}
The sensing performance using RIS can be measured by CRLB~\cite{6600800,9684752,9506874}. Collect all the unknown parameters into a vector
\begin{align}
\boldsymbol{\zeta} \triangleq \begin{bmatrix}
\boldsymbol{\theta}^{\text{T}}_{\text{TR}}, \boldsymbol{z}^{\text{T}}(t), q(t)		
	\end{bmatrix}^{\text{T}},
\end{align}
and the probability density function of the received signal can be expressed as
\begin{align}
f(\boldsymbol{r}(t);\boldsymbol{\zeta}) = \frac{1}{\pi^N\det(\boldsymbol{\Sigma})}e^{-[\boldsymbol{r}(t)-\boldsymbol{\mu}(t)]^{\text{H}}\boldsymbol{\Sigma}^{-1}[\boldsymbol{r}(t)-\boldsymbol{\mu}(t)]},
\end{align}
where the mean $\boldsymbol{\mu}$ and covariance matrix $\boldsymbol{\Sigma}$ are respectively
\begin{align}
\boldsymbol{\mu}	&=\boldsymbol{G}\boldsymbol{A}(\boldsymbol{\theta}_{\text{TR}})\boldsymbol{z}(t)	+\boldsymbol{G}\boldsymbol{a}(\theta_{\text{AR}})q(t),\\
\boldsymbol{\Sigma} &= \sigma^2_{\text{w}}\boldsymbol{I}_N.
\end{align}
We denote the noise variance as $\sigma^2_{\text{w}}$.

To obtain the CRLB, the Fisher information matrix (FIM) can be calculated first, defined as $\boldsymbol{F}$. With the likelihood function $\ln f(\boldsymbol{r}(t); \boldsymbol{\zeta})$, the FIM can be obtained as
\begin{align}
\boldsymbol{F} = 
\begin{bmatrix}
\boldsymbol{\Omega}_{1,1} & \boldsymbol{\Omega}_{1,2} & \boldsymbol{\Omega}_{1,3} \\
 \boldsymbol{\Omega}_{2,1} & \boldsymbol{\Omega}_{2,2} & \boldsymbol{\Omega}_{2,3} \\
 \boldsymbol{\Omega}_{3,1} & \boldsymbol{\Omega}_{3,2} & \boldsymbol{\Omega}_{3,3} 
\end{bmatrix}.	
\end{align}  
The expressions are given in Appendix~\ref{ap1}. Finally, the expression for FIM can be expressed as (\ref{fim}).
 
\begin{figure*}
	\normalsize
	\vspace*{4pt}
\begin{align}\label{fim}
\boldsymbol{F} = \sigma^{-2}_{\text{w}} \begin{bmatrix}
 2\mathcal{R}\{\boldsymbol{B}^{\text{H}}\boldsymbol{G}^{\text{H}}\boldsymbol{GB}\} & 	 \boldsymbol{B}^{\text{H}}\boldsymbol{G}^{\text{H}} \boldsymbol{G}\boldsymbol{A}(\boldsymbol{\theta}_{\text{TR}}) & 	\boldsymbol{B}^{\text{H}}\boldsymbol{G}^{\text{H}} \boldsymbol{G}\boldsymbol{a}(\theta_{\text{AR}})\\
 \boldsymbol{A}^{\text{H}}(\boldsymbol{\theta}_{\text{TR}})
\boldsymbol{G}^{\text{H}}  \boldsymbol{GB} &   \boldsymbol{A}^{\text{H}}(\boldsymbol{\theta}_{\text{TR}})
\boldsymbol{G}^{\text{H}}
\boldsymbol{G}\boldsymbol{A}(\boldsymbol{\theta}_{\text{TR}}) &  \boldsymbol{A}^{\text{H}}(\boldsymbol{\theta}_{\text{TR}})
\boldsymbol{G}^{\text{H}}  \boldsymbol{G}\boldsymbol{a}(\theta_{\text{AR}})\\
\boldsymbol{a}^{\text{H}}(\theta_{\text{AR}}) \boldsymbol{G}^{\text{H}} \boldsymbol{GB} &\boldsymbol{a}^{\text{H}}(\theta_{\text{AR}})
\boldsymbol{G}^{\text{H}}  \boldsymbol{G}
\boldsymbol{A}(\boldsymbol{\theta}_{\text{TR}}) & 
\boldsymbol{a}^{\text{H}}(\theta_{\text{AR}})\boldsymbol{G}^{\text{H}} \boldsymbol{G}\boldsymbol{a}(\theta_{\text{AR}})
\end{bmatrix}.
\end{align} 
		\hrulefill
\end{figure*}

The CRLB for the DOA estimation of the $k$-th vehicle can be expressed as
\begin{align}
f_\text{CRLB}(\theta_{\text{TR},k}) \geq [\boldsymbol{F}^{-1}]_{k,k}.
\end{align}
where $[A]_{m,n}$ denotes an entry of $\boldsymbol{A}$ at the $m$-th row and $n$-th column.

We give a simple example of CRLB for only one vehicle with the DOA being $\psi$. and the CRLB can be expressed as
\begin{align}
f_{\text{CRLB}}(\psi)	 & \geq [\boldsymbol{F}^{-1}]_{0,0}\geq \frac{\sigma^2_{\text{w}}}{ 2\mathcal{R}\{\boldsymbol{B}^{\text{H}}\boldsymbol{G}^{\text{H}}\boldsymbol{GB}\}}\\
& = \frac{\sigma^2_{\text{w}}}{ 2\boldsymbol{b}^{\text{H}}(\psi)\boldsymbol{G}^{\text{H}}\boldsymbol{Gb}(\psi)} \notag\\
& = \frac{\sigma^2_{\text{w}}}{ 2|z(t)|^2\nabla_{\psi}\boldsymbol{a}^{\text{H}} (\psi)\boldsymbol{G}^{\text{H}}\boldsymbol{G}\nabla_{\psi}\boldsymbol{a}(\psi)} \notag\\
& = \frac{\sigma^2_{\text{w}}d^2_{\text{AT}}d^2_{\text{TR}}d^2_{\text{RS}}}{2 |\gamma\alpha|^2 P_{\text{s}} } \left\|\boldsymbol{G}\nabla_{\psi}\boldsymbol{a}(\psi)\right\|^{-2}_2\notag
\end{align}
where $P_s$ is the power of the transmitted signal, $\alpha$ is determined by the vehicle scattering coefficient, and $\gamma$ is determined by the antenna gains of RIS and sensor. $d_{\text{AT}}$, $d_{\text{TR}}$ and $d_{\text{RS}}$ are the distance between the AP and the vehicle, the distance between the vehicle and the RIS, and the distance between the RIS and the sensor, respectively. 

\begin{figure}
	\centering
	\includegraphics[width=3in]{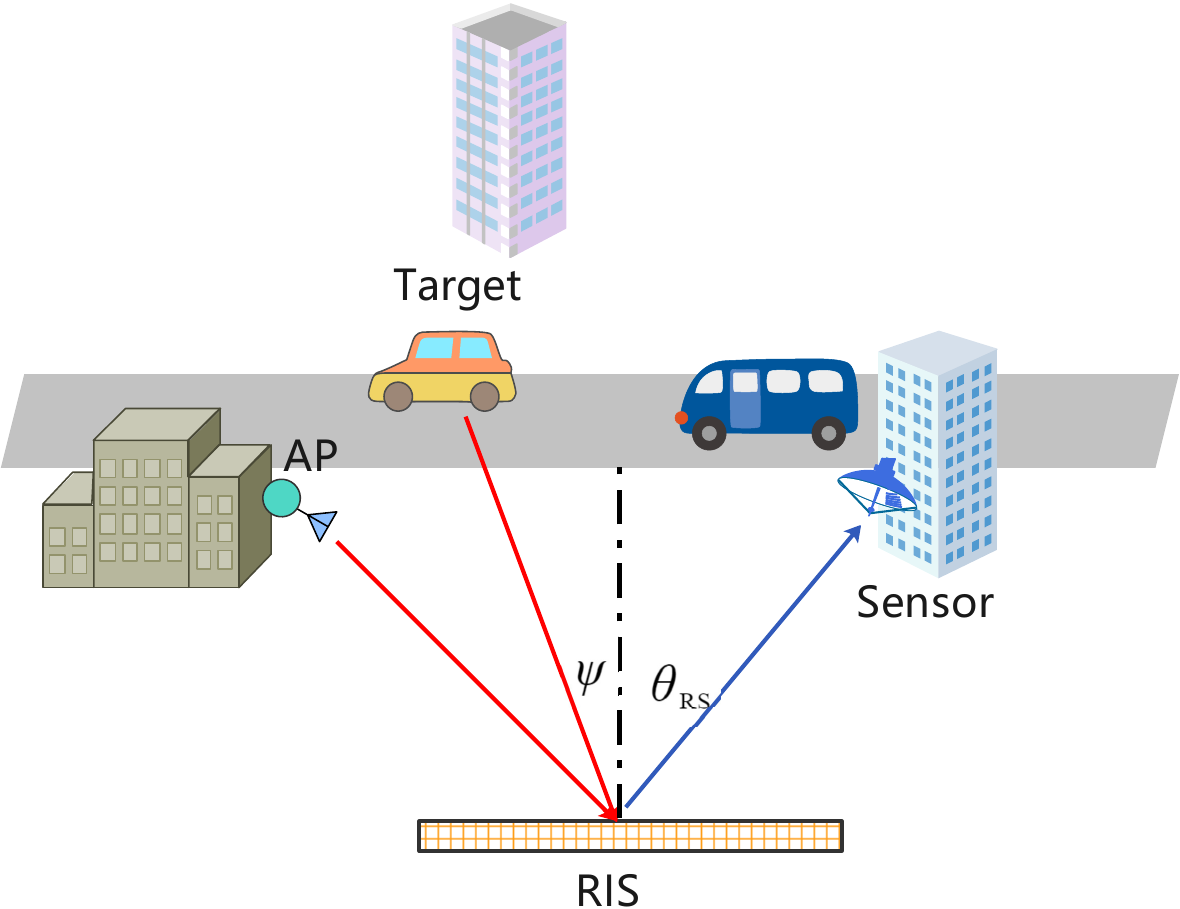}
	\caption{The diagram to optimize the measurement matrix}
	\label{meamat}
\end{figure}  

Therefore, we can decrease the CRLB to improve the DOA estimation performance. The following methods can be done to improve the performance:
\begin{enumerate}
	\item Decreasing the distance $d_{\text{AT}}$ between the AP and vehicle, the distance $d_{\text{TR}}$ between the vehicle and the RIS, and the distance $d_{\text{RS}}$ between the RIS and the sensor;
	\item Increasing the antenna gains of RIS and sensor;
	\item Increasing ratio $P_s/\sigma^2_{\text{w}}$ between the transmitting power and noise;
	\item Optimizing the measurement matrix $\boldsymbol{G}$ to improve $\left\|\boldsymbol{G}\nabla_{\psi}\boldsymbol{a}(\psi)\right\|^{2}_2$;
	\item Using the data compression method to improve the data sharing efficiency, especially in the vehicle applications~\cite{rs15082165}.
\end{enumerate}
For practical consideration, we can design the system by putting the sensor close to the RIS to decrease $d_{\text{RS}}$ and optimize the measurement matrix $\boldsymbol{G}$. As shown in Fig.~\ref{meamat}, we can optimize the measurement matrix $\boldsymbol{G}$ correspond to the steering vector of the angle $\psi+\theta_{\text{RS}}$. For convenience, we can set $\theta_{\text{RS}}=0$ to simplify the optimization of $\boldsymbol{G}$.

\section{Simulation Results}\label{simulation}
In this section, simulation results for the proposed passive sensing in the system using the RIS are carried out in a computer with Intel Core i5 @ 2.9 GHz processor, 8 GB DDR and MATLAB R2020b. Additionally, the code about the proposed method is available online \url{https://github.com/chenpengseu/PassiveDOA-ISAC-RIS.git}. The simulation parameters are given in Table~\ref{table1}. For the wireless communication parameters, we set the carrier frequency being $2.4$~GHz, the number of subcarriers being $64$, and the bandwidth being $20$~MHz. Moreover, we also show the performance with methods with and without the grids. In the methods with grids, the spatial domain is discretized into grids with the grid size being $\ang{1}$, and the methods without grids have a grid size being $\ang{0.01}$.  The parameters are chosen according to the practical vehicular sensing and communication applications. For example,   the distance between adjacent is usually $0.5$ wavelength in an array, the distances between the AP, RIS, sensor, and vehicles are chosen according to the width of a road, and the sensing range is chosen according to the array gain.

\begin{table}[!t]
	\renewcommand{\arraystretch}{1.3}
	\caption{Simulation Parameters}
	\label{table1}
	\centering
	\begin{tabular}{cc}
		\hline
		\textbf{Parameter} & \textbf{Value}\\
		\hline
		The direction between the RIS and  sensor & $\theta_{\text{Rs}}=0$\\
		The distance between adjacent elements & $d_{\text{E}}=0.5\lambda$\\
		The number of RIS elements & $M = 64$\\
		The direction between the AP and  RIS & $\theta_{\text{AR}}=\ang{-9.3878}$\\
		The number of vehicles & $K=3$\\
		The distance between the AP and  vehicles & $d_{\text{AT}}=20$ m\\
		The distance between the RIS and  vehicles  & $d_{\text{TR}}=30$ m \\
		The distance between the RIS and  sensor & $d_{\text{RS}}=3$ m\\
		The distance between the AP and  RIS & $d_{\text{AR}}=5$ m\\
		The number of measurements & $N=16$\\
		The spatial range & $[\ang{-45},\ang{45}]$\\
		The mean value of DOA & $\mathcal{E}\{\boldsymbol{\theta}_{\text{TR}}\}=[\ang{-25},\ang{15},\ang{30}]$\\
	\hline
	\end{tabular}
\end{table}

\begin{figure*}  
\centering  
\subfigure[The sensor is at $(20\text{ m},0\text{ m})$]{\label{crlbsub1} \includegraphics[width=0.45\textwidth]{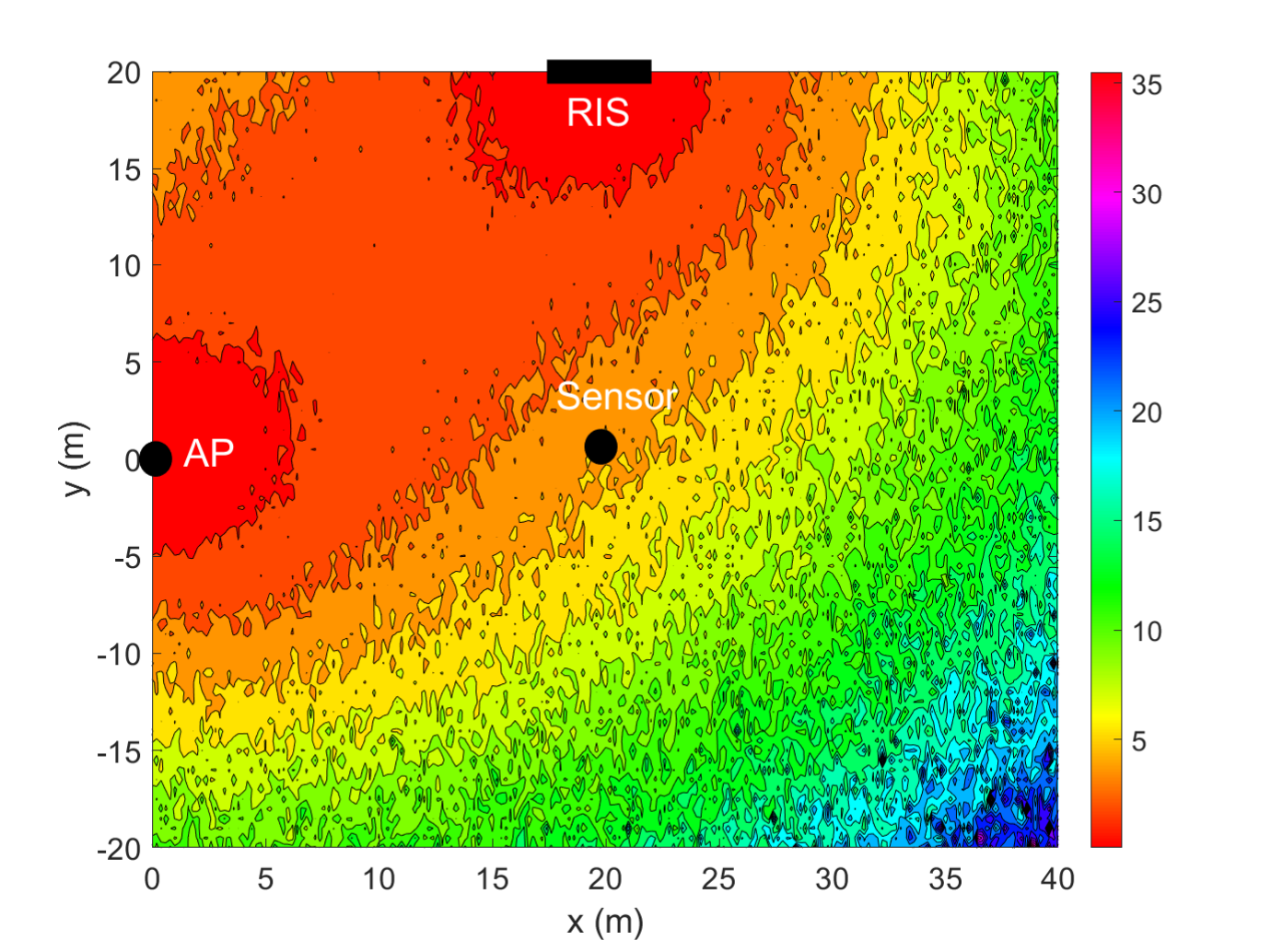}}
\subfigure[The sensor is at $(20\text{ m},17\text{ m})$]{\label{crlbsub2} \includegraphics[width=0.45\textwidth]{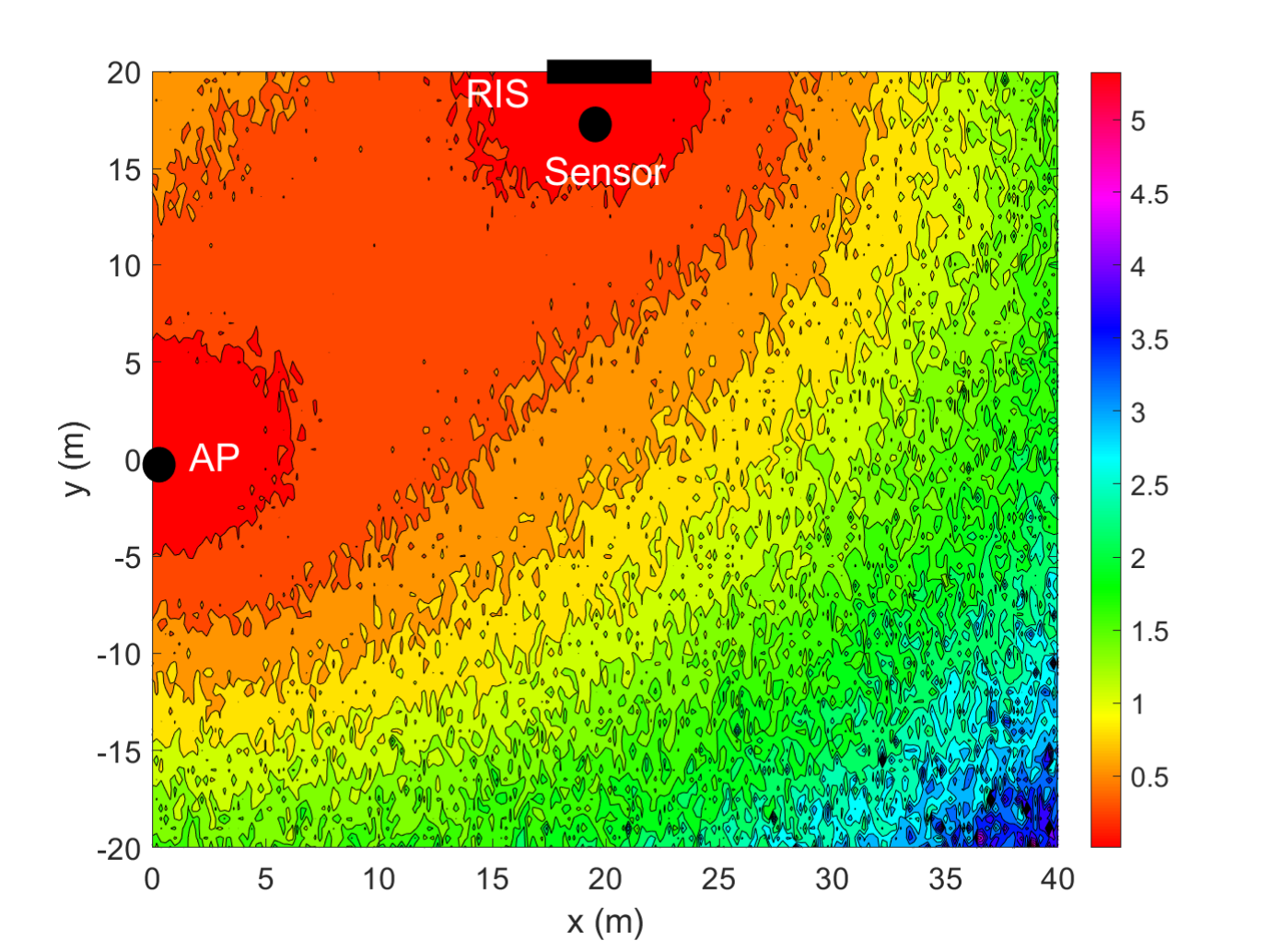}}
\subfigure[The AP is at $(20\text{ m},0\text{ m})$]{\label{crlbsub3} \includegraphics[width=0.45\textwidth]{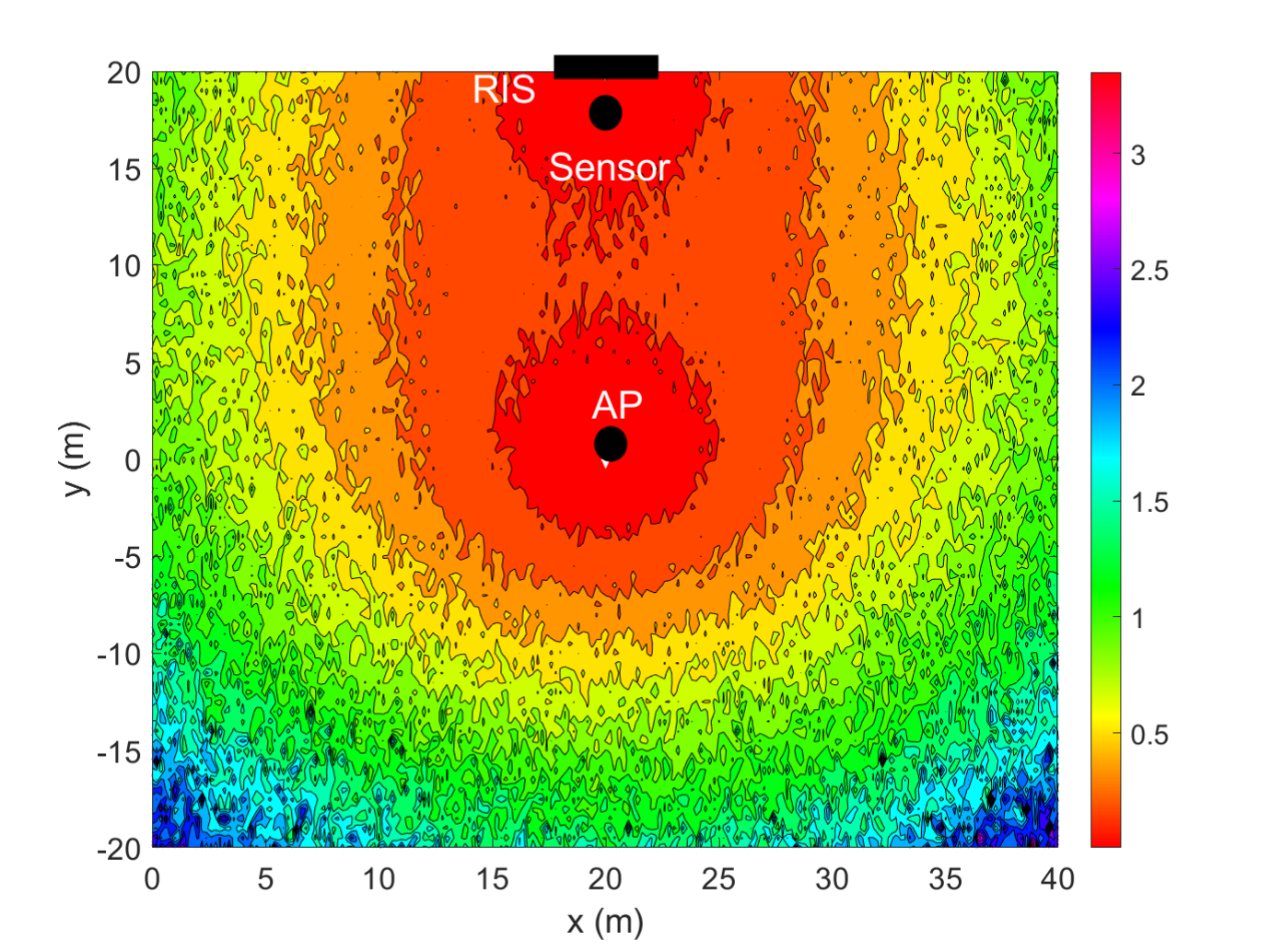}}
\subfigure[The AP is at $(20\text{ m},-20\text{ m})$]{\label{crlbsub4} \includegraphics[width=0.45\textwidth]{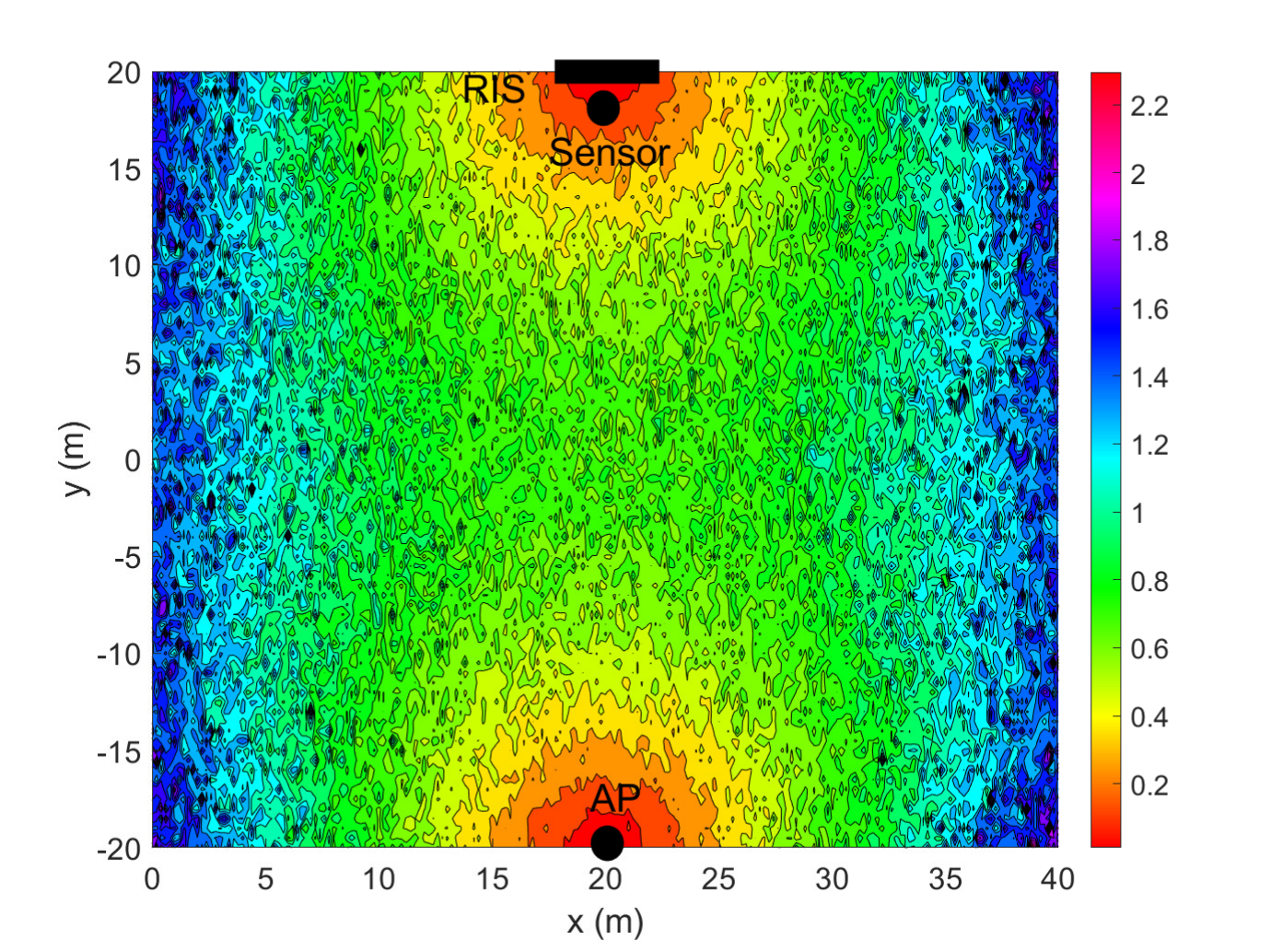}}
\caption{The RMSE (root of CRLB) with different positions of the AP and the sensor.}\label{crlball}
\end{figure*}

First, we show the root mean squared error (RMSE, the root of CRLB root) of the DOA estimation performance in the scenarios with different positions of the AP and the sensor, where the noise variance is $\sigma^2_{\text{w}}=0.01$. In~Fig.\ref{crlbsub1}, the AP is at $(0\text{ m},0\text{ m})$, the RIS is at $(20\text{ m},20\text{ m})$, and the sensor is at $(20\text{ m},0\text{ m})$. The different colors show the low bound of the DOA estimation in RMSE, and we can find that the area close to the AP and the RIS performs better than others. The worse RMSE is about $\ang{25}$ in the interesting area. Additionally, we also show the estimation low bound in the scenario with the sensor at $(20\text{ m},17\text{ m})$ in Fig.~\ref{crlbsub2}, and we can find that the overall RMSE is less than $\ang{4}$. Hence, better performance can be achieved by Fig.~\ref{crlbsub2} than by Fig.~\ref{crlbsub1}. Moreover, we move the AP to the position $(20\text{ m},0\text{ m})$ in Fig.~\ref{crlbsub3}, and move the AP to the position $(20\text{ m}, -20\text{ m})$ in Fig.~\ref{crlbsub4}. The area with better estimation performance is also moving with the AP. Fig.~\ref{crlbsub4} achieves the best performance than other $3$ scenarios. From the results in Fig.~\ref{crlball}, we can find that a better estimation performance can be achieved by putting the sensor close to the RIS.

\begin{figure}
	\centering
	\includegraphics[width=3.5in]{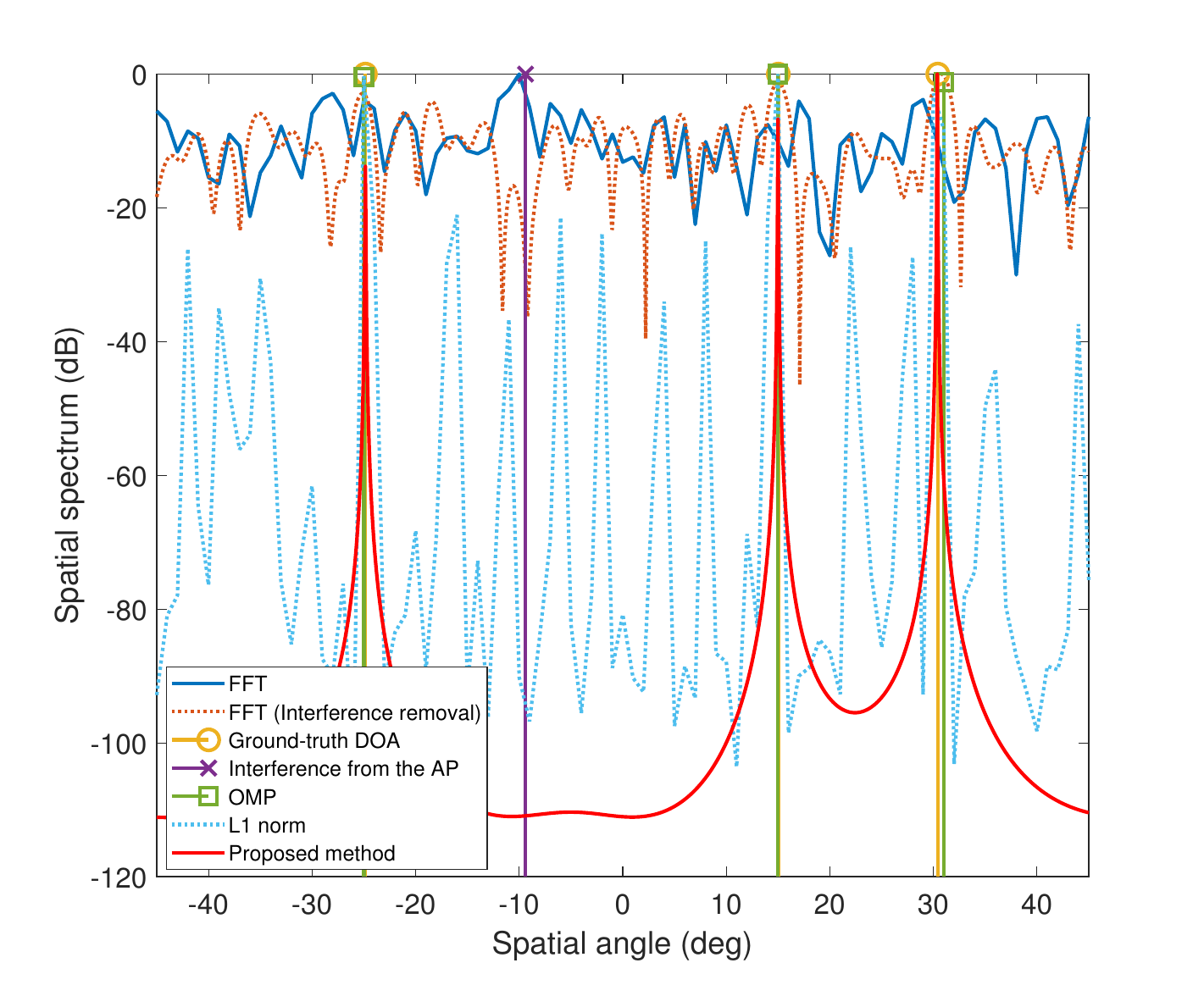}
	\caption{The estimated spatial spectrum with SNR being $20$ dB.}
	\label{sp}
\end{figure}  

Second, for the DOA estimation, we show the estimated spatial spectrum in Fig.~\ref{sp}, and we also compare it with the following methods:
\begin{itemize}
    \item FFT method: The traditional fast Fourier transformation (FFT) method estimates the spatial spectrum without discretizing the spatial domain, and the Rayleigh criterion limits the resolution. 
    \item FFT (interference removal) method: In this paper, the FFT method is used to estimate the DOA of vehicles after removing the interference from the AP. 
    \item OMP method~\cite{9320601,9328485,9409636}: 
    After removing the interference from the AP, an OMP method is used for the sparse reconstruction with lower computational complexity in the scenario with fewer grids but higher computational complexity with more grids. The matrix inverse step determines the computation complexity of the OMP method.
    \item $\ell_1$ norm method~\cite{6705656,7605512}: The $\ell_1$ norm minimization method is a convex optimization method for the sparse reconstruction, and we formulate the $\ell_1$ norm method as follows:
    \begin{equation}
        \begin{split}
            \min_{\Tilde{\boldsymbol{x}},\Tilde{q}}\,\left\|\boldsymbol{r}- \boldsymbol{G}\Tilde{\boldsymbol{D}}\boldsymbol{\Tilde{x}}	-\boldsymbol{G}\boldsymbol{a}(\theta_{\text{AR}})\Tilde{q}\right\|^2_2+\Tilde{\rho}\|\Tilde{\boldsymbol{x}}\|_1,
        \end{split}
    \end{equation} 
    where $\Tilde{\rho}$ is a parameter to control the balance between the sparsity, and the reconstruction error, $\Tilde{\boldsymbol{x}}$ is the sparse spectrum, $\Tilde{q}$ is the interference from the AP, and $\Tilde{\boldsymbol{D}}$ denotes the Fourier dictionary matrix for the $\ell_1$ reconstruction. 
\end{itemize}
As shown in Fig.~\ref{sp}, with the SNR being $20$ dB, the FFT and the FFT (interference removal) methods cannot estimate the spatial spectrum well since the number of measurements $N$ is much less than the number of RIS elements $M$. The proposed method, the $\ell_1$ norm method, and the OMP method can estimate the DOA well. Additionally, the proposed method can achieve better estimation performance than other methods.

\begin{figure}
	\centering
	\includegraphics[width=3.5in]{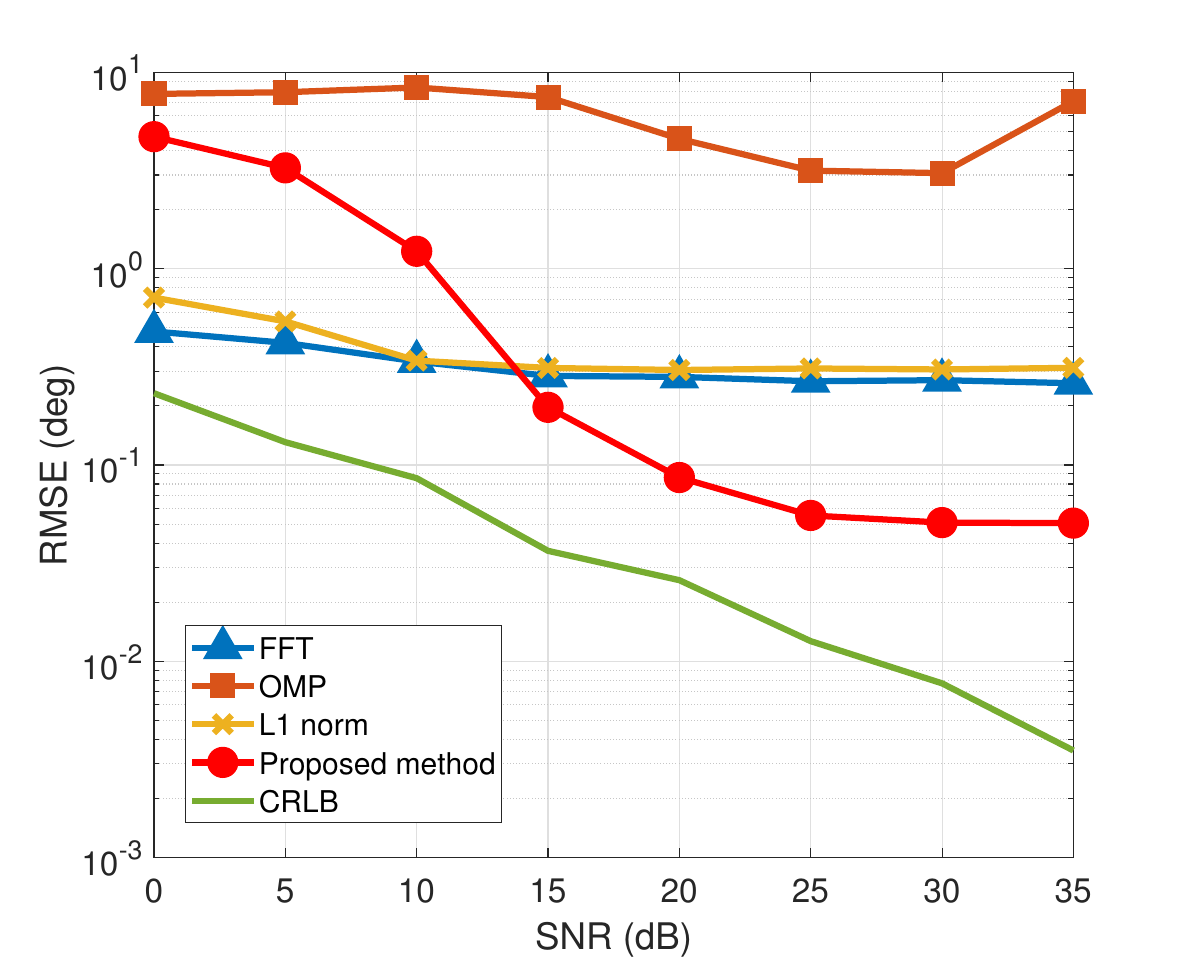}
	\caption{The estimated performance with different SNRs (without measurement matrix optimization).}
	\label{SNR}
\end{figure}

Then, we show the DOA estimation performance in the scenario with different SNRs in Fig.~\ref{SNR}, where the measurement matrix is chosen randomly and not optimized. We can find that the proposed method can outperform the existing methods when the SNR is greater than $15$~dB, where the RMSE is less than $\ang{0.2}$. However, with the improvement of SNR, the estimation performance cannot be improved when the SNR is greater than $25$~dB, where the RMSE is about $\ang{0.05}$. The RMSE platform is caused by interference from the AP.

\begin{figure}
	\centering
	\includegraphics[width=3.5in]{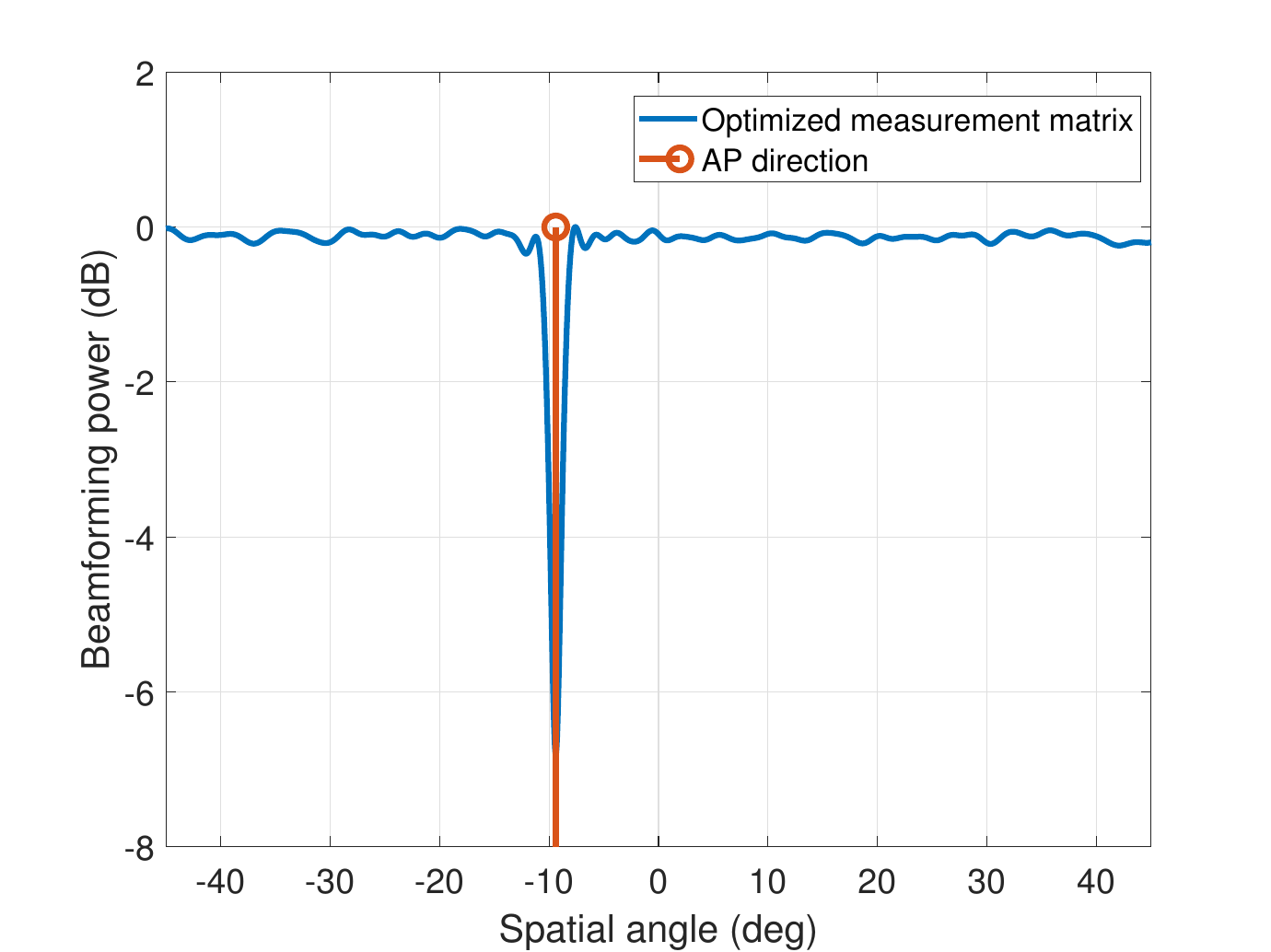}
	\caption{The beamforming results using the optimized measurement matrix.}
	\label{beam}
\end{figure} 

\begin{figure}
	\centering
	\includegraphics[width=3.5in]{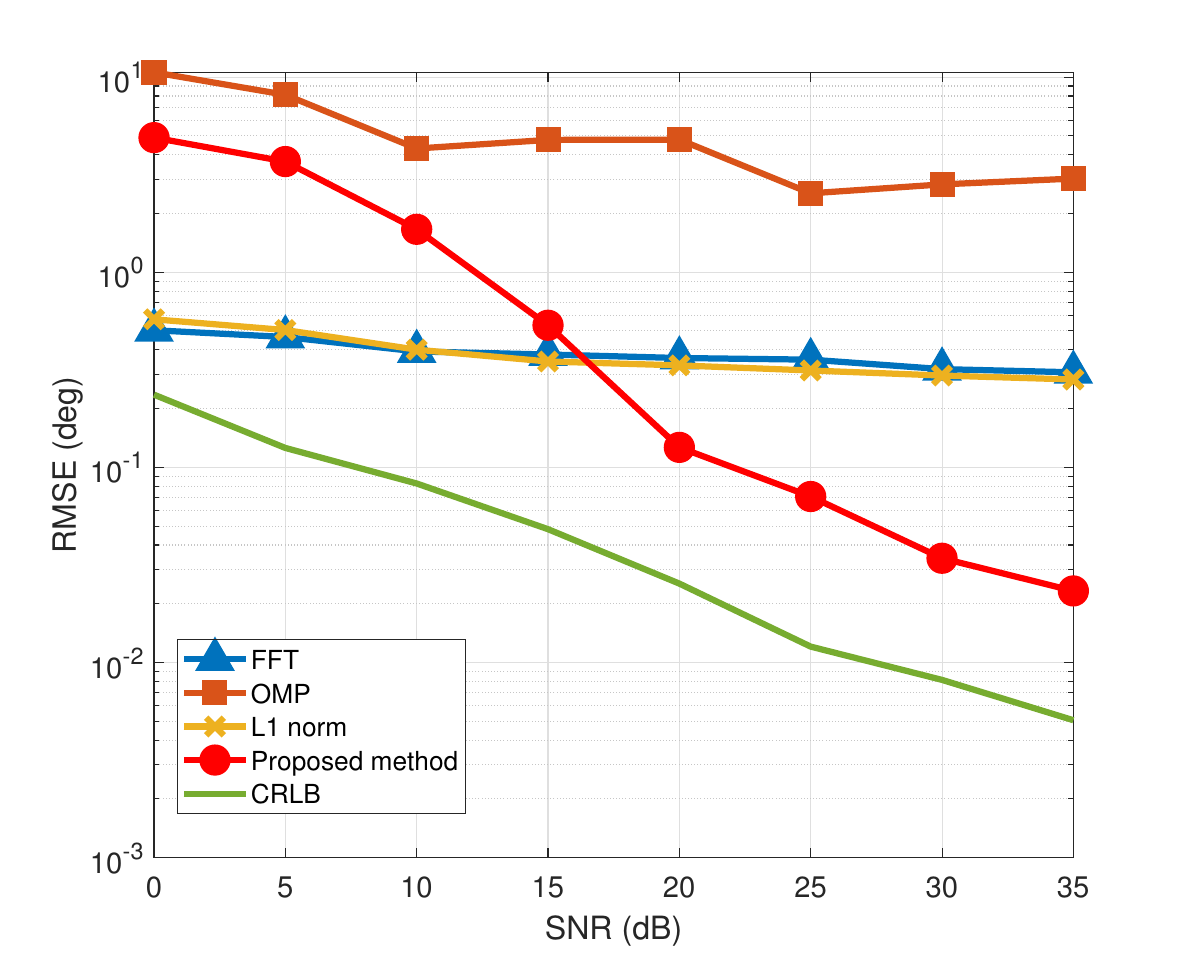}
	\caption{The estimated performance with different SNRs (with measurement matrix optimization).}
	\label{SNRwith}
\end{figure}

Moreover, we use the proposed method to optimize the measurement matrix. With the optimized matrix $\boldsymbol{G}$, the beamforming results are shown in Fig.~\ref{beam}. We can find that the beam power from the direction of the AP is much less than in other directions so that the interference can be removed significantly. Then, we use the optimized matrix $\boldsymbol{G}$ to perform the DOA estimation, and the estimation performance is shown in Fig.~\ref{SNRwith}. The better estimation performance is achieved in the scenario with $\text{SNR}\geq 20$~dB. Furthermore, the RMSE platform is also broken, and the performance can achieve the CRLB with the increasing SNR.

\begin{figure}
	\centering
	\includegraphics[width=3.5in]{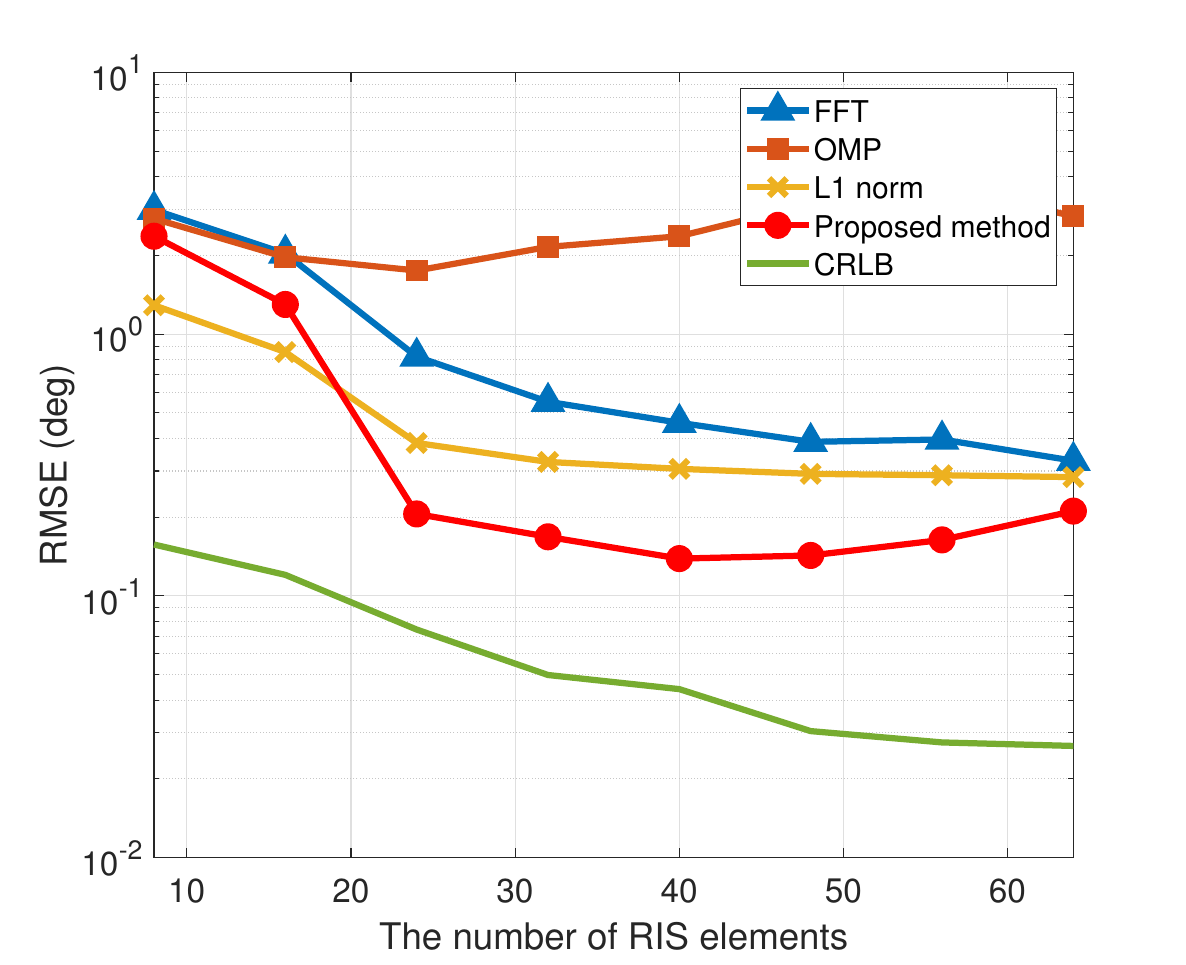}
	\caption{The estimation performance with different numbers of RIS elements (the SNR is $20$ dB).}
	\label{celements}
\end{figure}

\begin{figure}
	\centering
	\includegraphics[width=3.5in]{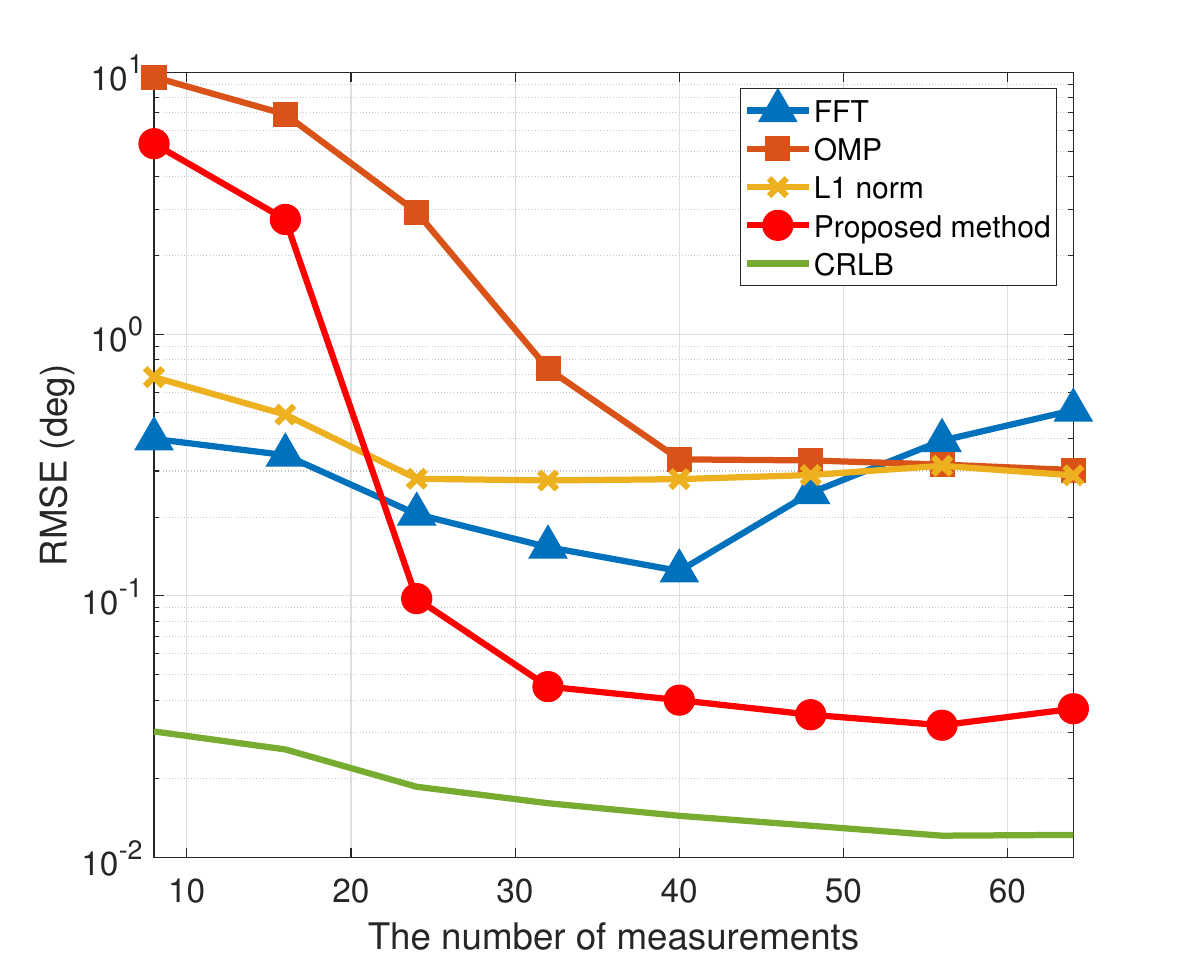}
	\caption{The estimation performance with different numbers of measurements (the SNR is $20$ dB).}
	\label{cmea}
\end{figure}

The DOA estimation performance with different numbers of RIS elements is shown in Fig.~\ref{celements}. We can find that the proposed method achieves a better estimation performance in the scenario with $M\geq 24$. However, the estimation performance cannot be further improved with more elements. Hence, in the proposed passive sensing system, the performance improvement with more elements is limited, and we can choose a suitable element number to satisfy the results.

\begin{figure}
	\centering
	\includegraphics[width=3.5in]{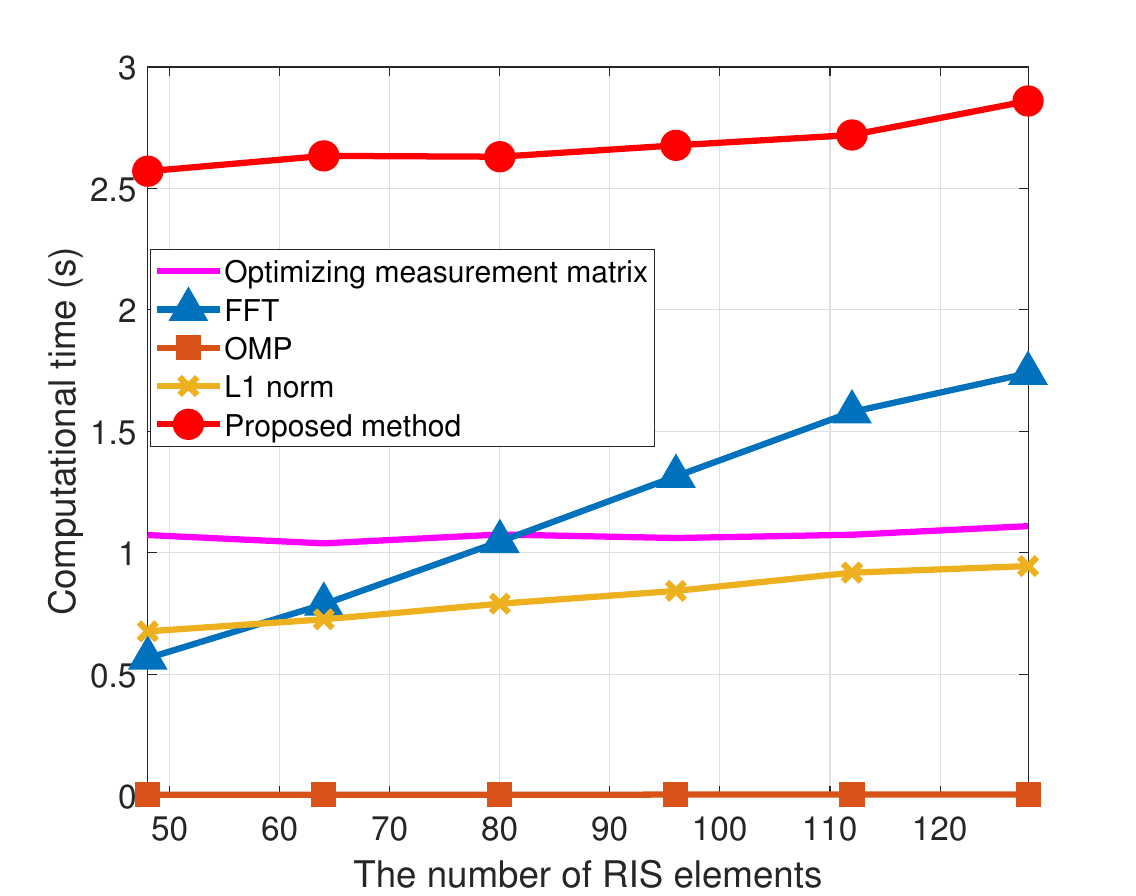}
	\caption{The computational time with different numbers of RIS elements.}
	\label{time}
\end{figure}

\begin{figure}
	\centering
	\includegraphics[width=3in]{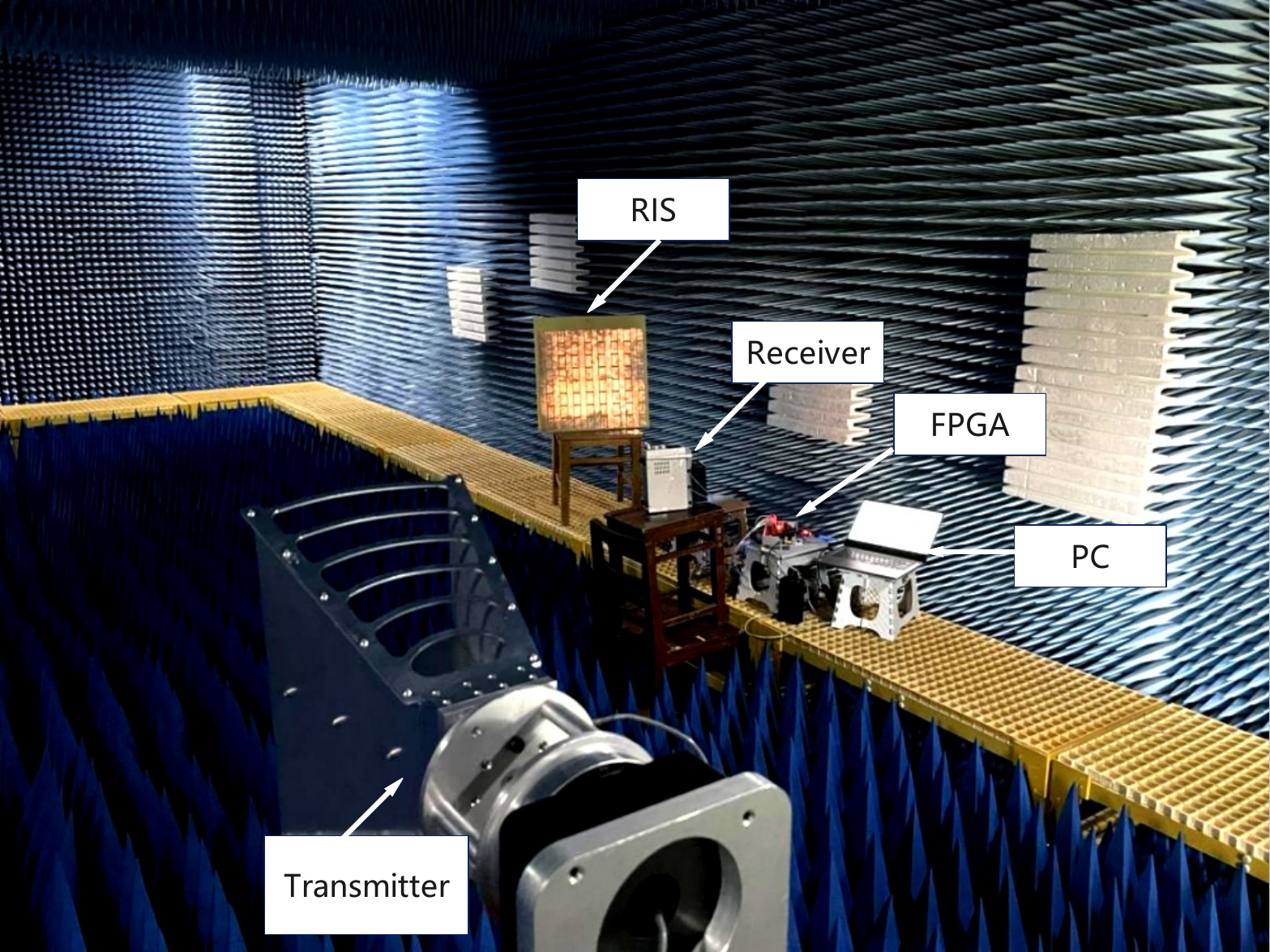}
	\caption{The RIS-based ISAC prototype system measured in the microwave anechoic chamber.}
	\label{pr}
\end{figure}

\begin{figure}
	\centering
	\includegraphics[width=3in]{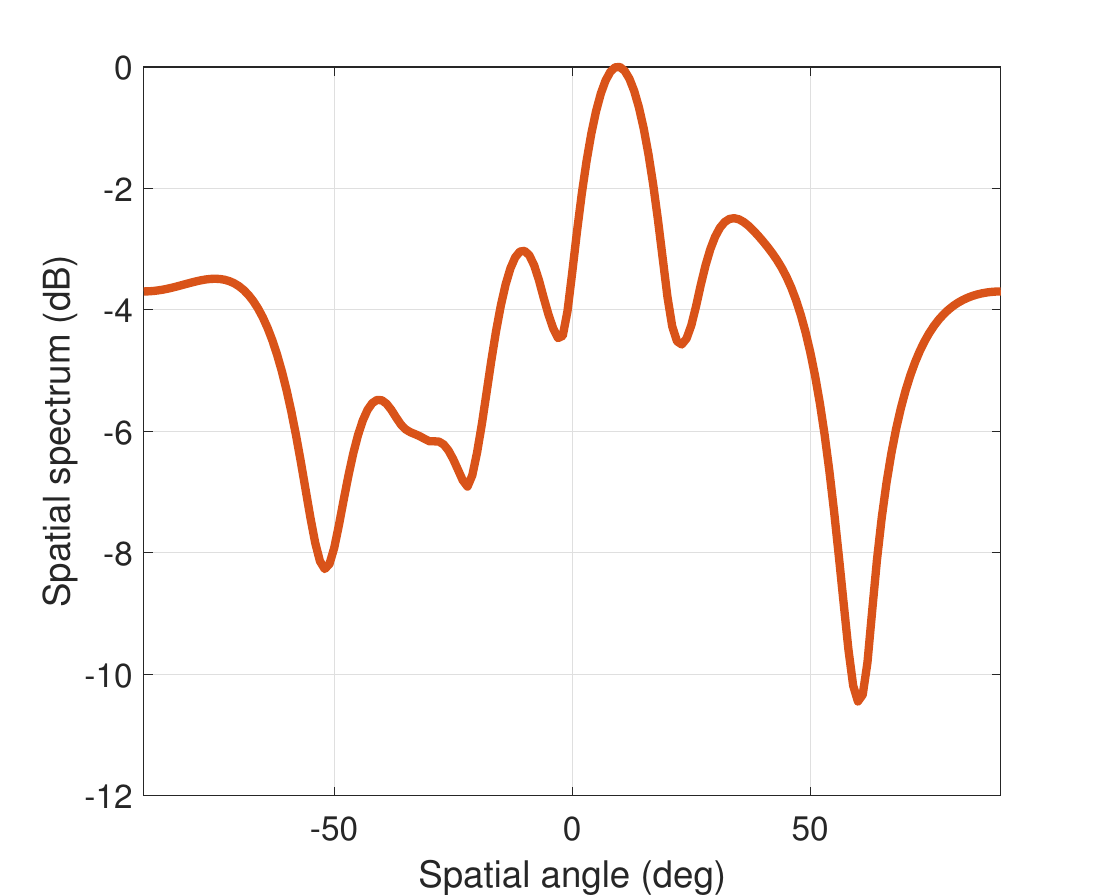}
	\caption{The measured spatial spectrum results of the RIS-based ISAC prototype system in microwave anechoic chamber.}
	\label{prmea}
\end{figure}

For the different $N$ measurements, the estimation performance is shown in Fig.~\ref{cmea}. We can see that more measurements can improve the estimation performance, and the performance of the proposed method cannot be further improved when $N\geq 32$, where the RMSE is about $\ang{0.04}$. For the OMP method, the RMSE platform is about $\ang{0.3}$. Hence, the proposed method is much better than the existing methods. Furthermore, we design and manufacture the RIS-based sensing system as shown in Fig.~\ref{pr}, where the system includes the transmitter, RIS, receiver, and field programmable gate array (FPGA) and personal computer (PC). The system is tested in a microwave anechoic chamber. The transmitter sends a $2.4$~GHz electromagnetic signals, and reflected by the RIS, which is controlled by  the FPGA. Then, the receiver samples the signals and transmits to the PC using TCP/IP ethernet protocol. The proposed algorithm is realized in the PC to reconstruct the spatial spectrum and estimate the DOA. The estimated spatial spectrum is shown in Fig.~\ref{prmea}, where we can get the target DOA by finding the peak value of the spatial spectrum. The results show that the proposed system can find the target direction efficiently. 

Finally, the computational time of the proposed method is shown in Fig.~\ref{time}. The computational time increases with more RIS elements when the proposed and FFT methods are adopted. The computational time of the proposed method is about $2.6$ s and is acceptable in most applications.  

\section{Conclusions}\label{conclusion}
The vehicle DOA estimation problem has been addressed in the proposed ISAC system with the RIS. The system model with the interference of wireless communication has been formulated. The atomic norm-based method has also been proposed for the vehicle DOA estimation with interference removal, and the spatial spectrum has been estimated by the Hankel-based MUSIC method. Additionally, the optimization problem has been solved for the measurement matrix to remove interference. The estimation performance has been measured by the CRLB, which can be used to optimize the distribution of the sensing node. The simulation results have shown the advantage of the proposed method in the DOA estimation. In the future work, considering the more powerful and efficient processors, we will focus on develop a more practical algorithm to improve the robust and improve the estimation performance, where more practical data will be collected to train and optimize the algorithm.
 
\bibliographystyle{IEEEtran}
\bibliography{IEEEabrv.bib,ref.bib} 

\appendices
\section{The Expression of FIM} \label{ap1}
With the likelihood function $\ln f(\boldsymbol{r}(t); \boldsymbol{\zeta})$, we have
\begin{equation}
	\begin{split}
		& \frac{\partial \ln f(\boldsymbol{r}(t); \boldsymbol{\zeta})}{\partial \boldsymbol{\theta}_{\text{TR}}}  =\frac{\partial\ln \frac{1}{\pi^N\det(\boldsymbol{\Sigma})}e^{-[\boldsymbol{r}(t)-\boldsymbol{\mu}(t)]^{\text{H}}\boldsymbol{\Sigma}^{-1}[\boldsymbol{r}(t)-\boldsymbol{\mu}(t)]}}{\partial\boldsymbol{\theta}_{\text{TR}}}\\
		&\qquad = -\frac{\partial [\boldsymbol{r}(t)-\boldsymbol{\mu}(t)]^{\text{H}}\boldsymbol{\Sigma}^{-1}[\boldsymbol{r}(t)-\boldsymbol{\mu}(t)]}{\partial\boldsymbol{\theta}_{\text{TR}}}\\
		&\qquad  =2\mathcal{R}\Bigg\{ [\boldsymbol{r}(t)-\boldsymbol{\mu}(t)]^{\text{H}} \boldsymbol{\Sigma}^{-1} \boldsymbol{G}\frac{\partial \boldsymbol{A}(\boldsymbol{\theta}_{\text{TR}})\boldsymbol{z}(t)}{\partial\boldsymbol{\theta}_{\text{TR}}}\Bigg\} \\	
		&\qquad  =2\mathcal{R}\Bigg\{ [\boldsymbol{r}(t)-\boldsymbol{\mu}(t)]^{\text{H}} \boldsymbol{\Sigma}^{-1} \boldsymbol{G}\boldsymbol{B}\Bigg\},   
	\end{split}    
\end{equation}
where the $k$-th ($k=0,1,\dots, K-1$) column of $\boldsymbol{B}$ is 
\begin{align}
	& \boldsymbol{b}(\theta_{\text{TR},k}) = z_k(t)
	\frac{\partial 
		\boldsymbol{a}(\theta_{\text{TR},k})}{\partial \theta_{\text{TR}},k}\\
	&\qquad = j\frac{2\pi d_{\text{E}}  z_k(t)}{\lambda}\cos(\theta_{\text{RS}}+\theta_{\text{TR},k})\left[\boldsymbol{a}(\theta_{\text{TR},k})\odot \boldsymbol{l}_M\right],\notag
\end{align}
and we define
\begin{align}
	\boldsymbol{l}_{M} = [0,1,\dots, M-1]^{\text{T}}.	
\end{align}

Additionally, we also have
\begin{align}
	& \frac{\partial \ln f(\boldsymbol{r}(t); \boldsymbol{\zeta})}{\partial \boldsymbol{z}(t)}  =\frac{\partial\ln \frac{1}{\pi^N\det(\boldsymbol{\Sigma})}e^{-[\boldsymbol{r}(t)-\boldsymbol{\mu}(t)]^{\text{H}}\boldsymbol{\Sigma}^{-1}[\boldsymbol{r}(t)-\boldsymbol{\mu}(t)]}}{\partial \boldsymbol{z}(t)}\notag\\
	&\qquad = -\frac{\partial [\boldsymbol{r}(t)-\boldsymbol{\mu}(t)]^{\text{H}}\boldsymbol{\Sigma}^{-1}[\boldsymbol{r}(t)-\boldsymbol{\mu}(t)]}{\partial \boldsymbol{z}(t)}\notag\\
	&\qquad = [\boldsymbol{r}(t)-\boldsymbol{\mu}(t)]^{\text{H}} \boldsymbol{\Sigma}^{-1} \frac{\partial \boldsymbol{G}\boldsymbol{A}(\boldsymbol{\theta}_{\text{TR}})\boldsymbol{z}(t)}{\partial \boldsymbol{z}(t)}\notag\\
	&\qquad\qquad +\left\{\boldsymbol{\Sigma}^{-1}[\boldsymbol{r}(t)-\boldsymbol{\mu}(t)]\right\}^{\text{T}}\frac{\partial \left[\boldsymbol{G}\boldsymbol{A}(\boldsymbol{\theta}_{\text{TR}})\boldsymbol{z}(t)\right]^*}{\partial \boldsymbol{z}(t)}\notag\\	
	&\qquad = [\boldsymbol{r}(t)-\boldsymbol{\mu}(t)]^{\text{H}} \boldsymbol{\Sigma}^{-1} \boldsymbol{G}\boldsymbol{A}(\boldsymbol{\theta}_{\text{TR}}). 
\end{align}

For the unknown signal $q(t)$, we have
\begin{align}
	& \frac{\partial \ln f(\boldsymbol{r}(t); \boldsymbol{\zeta})}{\partial q(t)}  =\frac{\partial\ln \frac{1}{\pi^N\det(\boldsymbol{\Sigma})}e^{-[\boldsymbol{r}(t)-\boldsymbol{\mu}(t)]^{\text{H}}\boldsymbol{\Sigma}^{-1}[\boldsymbol{r}(t)-\boldsymbol{\mu}(t)]}}{\partial q(t)}\notag\\
	&\qquad = -\frac{\partial [\boldsymbol{r}(t)-\boldsymbol{\mu}(t)]^{\text{H}}\boldsymbol{\Sigma}^{-1}[\boldsymbol{r}(t)-\boldsymbol{\mu}(t)]}{\partial q(t)}\notag\\
	&\qquad = [\boldsymbol{r}(t)-\boldsymbol{\mu}(t)]^{\text{H}} \boldsymbol{\Sigma}^{-1} \frac{\partial \boldsymbol{\mu}(t)}{\partial q(t)}\notag\\
	&\qquad\qquad +\left\{\boldsymbol{\Sigma}^{-1}[\boldsymbol{r}(t)-\boldsymbol{\mu}(t)]\right\}^{\text{T}}\frac{\partial \boldsymbol{\mu}^{\text{*}}(t)}{\partial q(t)}\notag\\
	&\qquad = [\boldsymbol{r}(t)-\boldsymbol{\mu}(t)]^{\text{H}} \boldsymbol{\Sigma}^{-1} \boldsymbol{G}\boldsymbol{a}(\theta_{\text{AR}}).
\end{align}

Therefore, the sub-matrices $\boldsymbol{\Omega}_{1,1}$, $\boldsymbol{\Omega}_{1,2}$, $\boldsymbol{\Omega}_{1,3}$, $\boldsymbol{\Omega}_{2,1}$, $\boldsymbol{\Omega}_{2,2}$, $\boldsymbol{\Omega}_{2,3}$, $\boldsymbol{\Omega}_{3,1}$, $\boldsymbol{\Omega}_{3,2}$, $\boldsymbol{\Omega}_{3,3}$ in FIM are obtained as follows:
\begin{itemize} 
\item $\boldsymbol{\Omega}_{1,1}$ is obtained as
\begin{align}
& \boldsymbol{\Omega}_{1,1}  = \mathcal{E}\left\{
\frac{\partial \ln^{\text{H}}f(\boldsymbol{r}(t); \boldsymbol{\zeta})}{\partial \boldsymbol{\theta}_{\text{TR}}}
\frac{\partial \ln f(\boldsymbol{r}(t); \boldsymbol{\zeta})}{\partial \boldsymbol{\theta}_{\text{TR}}}
\right\} \\
&\quad = \mathcal{E}\Bigg\{\sigma^{-2}_{\text{w}}\big\{\boldsymbol{B}^{\text{H}}\boldsymbol{G}^{\text{H}}[\boldsymbol{r}(t)-\boldsymbol{\mu}(t)]+\boldsymbol{B}^{\text{T}}\boldsymbol{G}^{\text{T}}\notag\\
&\qquad\quad [\boldsymbol{r}(t)-\boldsymbol{\mu}(t)]^{*}
\big\}\sigma^{-2}_{\text{w}}\big\{[\boldsymbol{r}(t)-\boldsymbol{\mu}(t)]^{\text{H}}\boldsymbol{GB}
\notag\\
&\qquad \quad +[\boldsymbol{r}(t)-\boldsymbol{\mu}(t)]^{\text{T}}\boldsymbol{G}^*\boldsymbol{B}^*\big\}\Bigg\}\notag\\
&\quad =  2\sigma^{-2}_{\text{w}}\mathcal{R}\{\boldsymbol{B}^{\text{H}}\boldsymbol{G}^{\text{H}}\boldsymbol{GB}\}. \notag
\end{align}
\item $\boldsymbol{\Omega}_{2,2}$ is obtained as
\begin{align}
& \boldsymbol{\Omega}_{2,2} = \mathcal{E}\left\{
\frac{\partial \ln^{\text{H}} f(\boldsymbol{r}(t); \boldsymbol{\zeta})}{\partial \boldsymbol{z}(t)}
\frac{\partial \ln f(\boldsymbol{r}(t); \boldsymbol{\zeta})}{\partial \boldsymbol{z}(t)}
\right\}\\
&\quad = \mathcal{E}\Bigg\{
\sigma^{-2}_{\text{w}} \boldsymbol{A}^{\text{H}}(\boldsymbol{\theta}_{\text{TR}})
\boldsymbol{G}^{\text{H}}
[\boldsymbol{r}(t)-\boldsymbol{\mu}(t)] \notag\\
&\qquad\quad
\sigma^{-2}_{\text{w}}
[\boldsymbol{r}(t)-\boldsymbol{\mu}(t)]^{\text{H}}  \boldsymbol{G}\boldsymbol{A}(\boldsymbol{\theta}_{\text{TR}})
\Bigg\}\notag\\
&\quad = 
\sigma^{-2}_{\text{w}} \boldsymbol{A}^{\text{H}}(\boldsymbol{\theta}_{\text{TR}})
\boldsymbol{G}^{\text{H}}
\boldsymbol{G}\boldsymbol{A}(\boldsymbol{\theta}_{\text{TR}}). \notag
\end{align}
\item $\boldsymbol{\Omega}_{3,3}$ is obtained as
\begin{align}
& \boldsymbol{\Omega}_{3,3} =\mathcal{E}\Bigg\{	\frac{\partial \ln^{\text{H}} f(\boldsymbol{r}(t); \boldsymbol{\zeta})}{\partial q(t)}\frac{\partial \ln f(\boldsymbol{r}(t); \boldsymbol{\zeta})}{\partial q(t)}\Bigg\}\\
&\quad = \mathcal{E}\Bigg\{
 \sigma^{-2}_{\text{w}}
\boldsymbol{a}^{\text{H}}(\theta_{\text{AR}})\boldsymbol{G}^{\text{H}}[\boldsymbol{r}(t)-\boldsymbol{\mu}(t)]\notag\\
&\qquad\quad
\sigma^{-2}_{\text{w}}
[\boldsymbol{r}(t)-\boldsymbol{\mu}(t)]^{\text{H}} \boldsymbol{G}\boldsymbol{a}(\theta_{\text{AR}})
\Bigg\}\notag\\
&\quad = \sigma^{-2}_{\text{w}}
\boldsymbol{a}^{\text{H}}(\theta_{\text{AR}})\boldsymbol{G}^{\text{H}} \boldsymbol{G}\boldsymbol{a}(\theta_{\text{AR}}).\notag
\end{align}

\item $\boldsymbol{\Omega}_{1,2}$ is obtained as
\begin{align}
& \boldsymbol{\Omega}_{1,2}  = 	\mathcal{E}\left\{
\frac{\partial \ln^{\text{H}}f(\boldsymbol{r}(t); \boldsymbol{\zeta})}{\partial \boldsymbol{\theta}_{\text{TR}}}
\frac{\partial \ln f(\boldsymbol{r}(t); \boldsymbol{\zeta})}{\partial \boldsymbol{z}(t)}
\right\} \\
&\quad = \mathcal{E}\Bigg\{\sigma^{-2}_{\text{w}}\big\{\boldsymbol{B}^{\text{H}}\boldsymbol{G}^{\text{H}}[\boldsymbol{r}(t)-\boldsymbol{\mu}(t)]+\boldsymbol{B}^{\text{T}}\boldsymbol{G}^{\text{T}}\notag\\
&\qquad\quad
[\boldsymbol{r}(t)-\boldsymbol{\mu}(t)]^{*}
\big\} \sigma^{-2}_{\text{w}}
[\boldsymbol{r}(t)-\boldsymbol{\mu}(t)]^{\text{H}}  \boldsymbol{G}\boldsymbol{A}(\boldsymbol{\theta}_{\text{TR}})
\Bigg\}\notag\\
&\quad =  \sigma^{-2}_{\text{w}} \boldsymbol{B}^{\text{H}}\boldsymbol{G}^{\text{H}} \boldsymbol{G}\boldsymbol{A}(\boldsymbol{\theta}_{\text{TR}}). \notag
\end{align}

\item $\boldsymbol{\Omega}_{1,3}$ is obtained as
\begin{align}
&\boldsymbol{\Omega}_{1,3}  = \mathcal{E}\Bigg\{
\frac{\partial \ln^{\text{H}}f(\boldsymbol{r}(t); \boldsymbol{\zeta})}{\partial \boldsymbol{\theta}_{\text{TR}}}
\frac{\partial \ln f(\boldsymbol{r}(t); \boldsymbol{\zeta})}{\partial q(t)}
\Bigg\}\\
&\quad = \mathcal{E}\Bigg\{ \sigma^{-2}_{\text{w}}\big\{\boldsymbol{B}^{\text{H}}\boldsymbol{G}^{\text{H}}[\boldsymbol{r}(t)-\boldsymbol{\mu}(t)]+\boldsymbol{B}^{\text{T}}\boldsymbol{G}^{\text{T}}\notag\\
&\qquad\quad[\boldsymbol{r}(t)-\boldsymbol{\mu}(t)]^{*}
\big\} \sigma^{-2}_{\text{w}}
[\boldsymbol{r}(t)-\boldsymbol{\mu}(t)]^{\text{H}} \boldsymbol{G}\boldsymbol{a}(\theta_{\text{AR}})
\Bigg\}\notag\\
&\quad = \sigma^{-2}_{\text{w}} \boldsymbol{B}^{\text{H}}\boldsymbol{G}^{\text{H}} \boldsymbol{G}\boldsymbol{a}(\theta_{\text{AR}}). \notag
\end{align}

\item $\boldsymbol{\Omega}_{2,1}$ is obtained as
\begin{align}
& \boldsymbol{\Omega}_{2,1}  =\mathcal{E}\left\{
\frac{\partial \ln^{\text{H}} f(\boldsymbol{r}(t); \boldsymbol{\zeta})}{\partial \boldsymbol{\theta}_{\text{TR}}}
\frac{\partial \ln f(\boldsymbol{r}(t); \boldsymbol{\zeta})}{\partial q(t)}
\right\}\\
&\quad= \sigma^{-2}_{\text{w}} \boldsymbol{A}^{\text{H}}(\boldsymbol{\theta}_{\text{TR}})
\boldsymbol{G}^{\text{H}}  \boldsymbol{GB}. \notag 
\end{align}

\item $\boldsymbol{\Omega}_{2,3}$ is obtained as
\begin{align}
& \boldsymbol{\Omega}_{2,3} =  \mathcal{E}\left\{
\frac{\partial \ln^{\text{H}} f(\boldsymbol{r}(t); \boldsymbol{\zeta})}{\partial \boldsymbol{z}(t)}
\frac{\partial \ln f(\boldsymbol{r}(t); \boldsymbol{\zeta})}{\partial q(t)}
\right\}\\
&\quad=\sigma^{-2}_{\text{w}} \boldsymbol{A}^{\text{H}}(\boldsymbol{\theta}_{\text{TR}})
\boldsymbol{G}^{\text{H}}  \boldsymbol{G}\boldsymbol{a}(\theta_{\text{AR}}).
\notag
\end{align}

\item $\boldsymbol{\Omega}_{3,1}$ is obtained as
\begin{align}
& \boldsymbol{\Omega}_{3,1} = \mathcal{E}\Bigg\{
\frac{\partial \ln^{\text{H}} f(\boldsymbol{r}(t); \boldsymbol{\zeta})}{\partial q(t)}
\frac{\partial \ln f(\boldsymbol{r}(t); \boldsymbol{\zeta})}{\partial \boldsymbol{\theta}_{\text{TR}}}
\Bigg\}\\
&\quad = \sigma^{-2}_{\text{w}} \boldsymbol{a}^{\text{H}}(\theta_{\text{AR}}) \boldsymbol{G}^{\text{H}} \boldsymbol{GB}.\notag
\end{align}

\item $\boldsymbol{\Omega}_{3,2}$ is obtained as
\begin{align}
&\boldsymbol{\Omega}_{3,2}  =  \mathcal{E}\left\{
\frac{\partial \ln^{\text{H}} f(\boldsymbol{r}(t); \boldsymbol{\zeta})}{\partial q(t)}
\frac{\partial \ln f(\boldsymbol{r}(t); \boldsymbol{\zeta})}{\partial \boldsymbol{z}(t)}
\right\}\\
&\quad=\sigma^{-2}_{\text{w}} \boldsymbol{a}^{\text{H}}(\theta_{\text{AR}})
\boldsymbol{G}^{\text{H}}  \boldsymbol{G}
\boldsymbol{A}(\boldsymbol{\theta}_{\text{TR}}).
\notag
\end{align}
\end{itemize}

\begin{IEEEbiography}[{\includegraphics[width=1in,height=1.25in,clip,keepaspectratio]{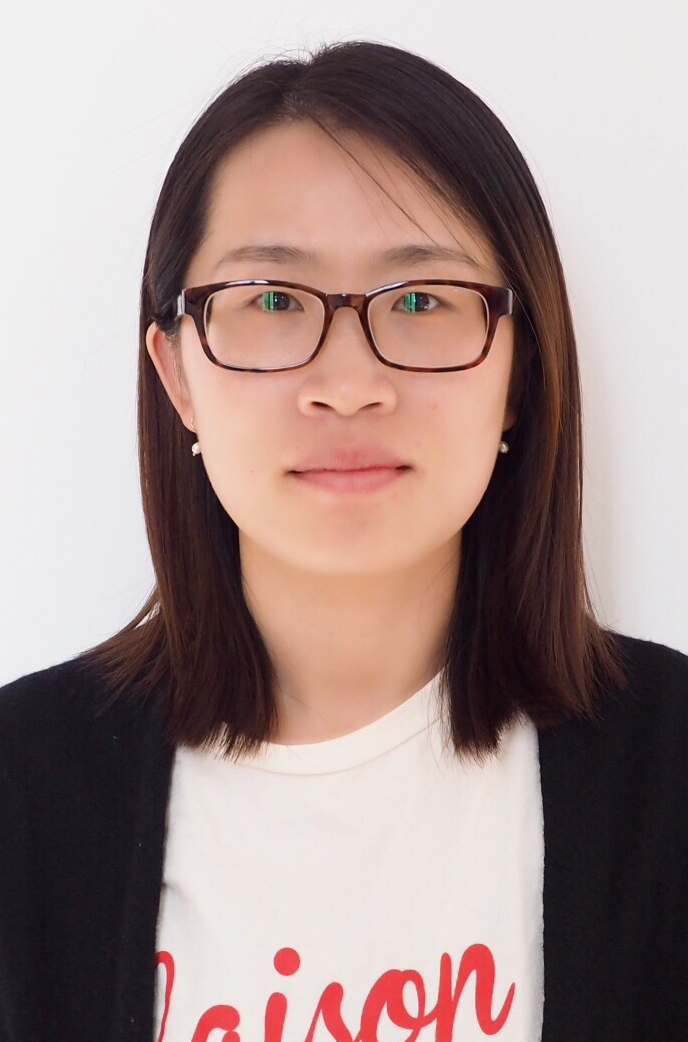}}]{Zhimin Chen (Member, IEEE)} received the Ph.D. degree in information and communication engineering from the School of Information Science and Engineering, Southeast University, Nanjing, China in 2015. Since 2015, she is currently an associate professor at Shanghai Dianji University, Shanghai, China. From 2021,  she is also a Visiting Scholar in  the Department of Electronic and Information Engineering, The Hong Kong Polytechnic University, Hong Kong.  Her research interests include array signal processing, vehicle communications and millimeter-wave communications. 
\end{IEEEbiography}

\begin{IEEEbiography}[{\includegraphics[width=1in,height=1.25in,clip,keepaspectratio]{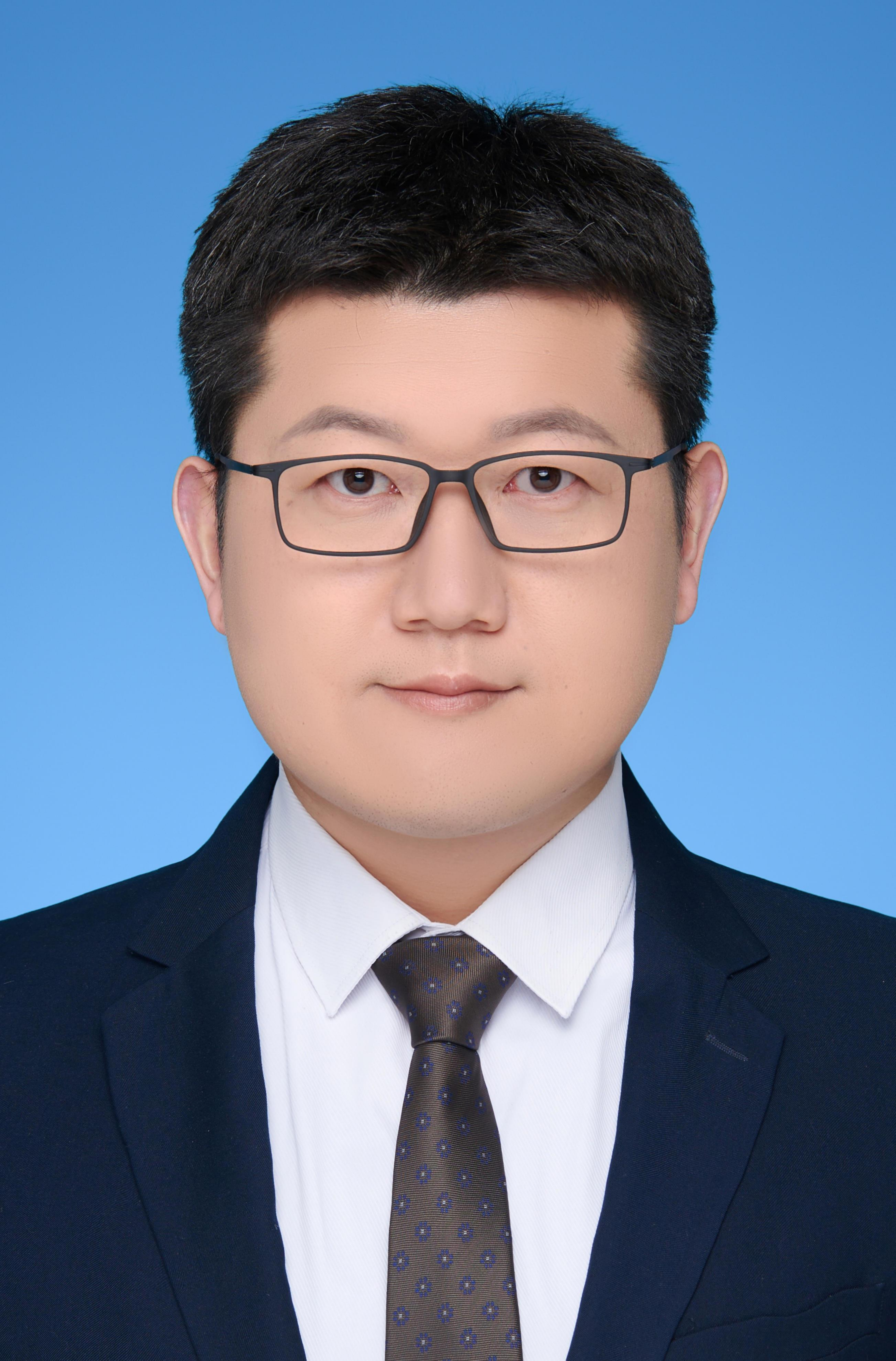}}]{Peng Chen (Seinor Member, IEEE)} received the B.E. and Ph.D. degrees from the School of Information Science and Engineering, Southeast University, Nanjing, China, in 2011 and 2017 respectively. From March 2015 to April 2016, he was a Visiting Scholar with the Department of Electrical Engineering, Columbia University, New York, NY, USA. He is currently an Associate Professor with the State Key Laboratory of Millimeter Waves, Southeast University. His research interests include target localization, super-resolution reconstruction, and array signal processing. He is a Jiangsu Province Outstanding Young Scientist. He has served as an IEEE ICCC Session Chair, and won the Best Presentation Award in 2022 (IEEE ICCC). He was invited as a keynote speaker at the IEEE ICET in 2022. He was recognized as an exemplary reviewer for IEEE WCL in 2021, and won the Best Paper Award at IEEE ICCCCEE in 2017.
\end{IEEEbiography}

 \begin{IEEEbiography}[{\includegraphics[width=1in,height=1.25in,clip,keepaspectratio]{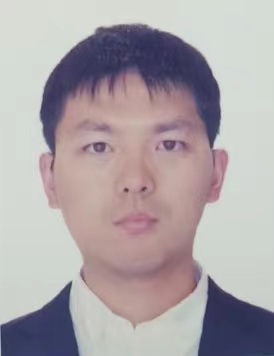}}]{Ziyu Guo (Member, IEEE)} received the B.E. and Ph.D. degrees from the School of Information Science and Engineering, Southeast University, Nanjing, China, in 2011 and 2016, respectively. From 2013 to 2015, he was a Visiting Student with Columbia University, New York, NY, USA. From 2019 to 2022, he was a Post-Doctoral Researcher with the State Key Laboratory of ASIC and System, Fudan University, Shanghai, China. He is currently an Assistant Professor of the School of Information Science and Technology, Fudan University. His research interests include millimeter-wave communication and VLSI design for digital signal processing.
 \end{IEEEbiography}
 
\begin{IEEEbiography}[{\includegraphics[width=1in,height=1.25in,clip,keepaspectratio]{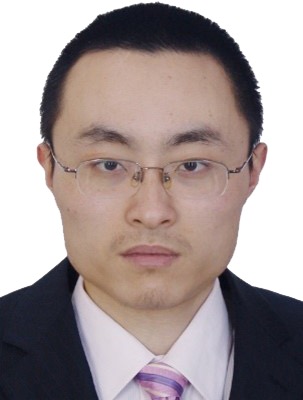}}]{Yudong Zhang (Seinor Member, IEEE)} received a Ph.D. in Signal and Information Processing from Southeast University in 2010. He worked as a postdoc from 2010 to 2012 with Columbia University, USA, and as an Assistant Research Scientist from 2012 to 2013 with the Research Foundation of Mental Hygiene (RFMH), USA. He served as a Full Professor from 2013 to 2017 at Nanjing Normal University. He serves as a Chair Professor from Dec/2017 at the School of Computing and Mathematical Sciences, University of Leicester, UK.
\end{IEEEbiography}

\begin{IEEEbiography}
	[{\includegraphics[width=1in,height=1.25in,clip,keepaspectratio]{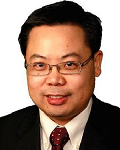}}]{Xianbin Wang (Fellow, IEEE)} received the Ph.D. degree in electrical and computer engineering from National University of Singapore, in 2001. He is a Professor and Tier-I Canada Research Chair at Western University, Canada. Prior to joining Western University, he was with Communications Research Centre (CRC) Canada as a Research Scientist/Senior Research Scientist between July 2002 and December 2007. From January 2001 to July 2002, he was a System Designer with STMicroelectronics, where he was responsible for the system design of DSL and Gigabit Ethernet chipsets. His current research interests include 5G technologies, Internet of Things, communications security, machine learning and locationing technologies. Dr. Wang has over 300 peer-reviewed journal and conference papers, in addition to 26 granted and pending patents and several standard contributions. Dr. Wang is a Fellow of Canadian Academy of Engineering and an IEEE Distinguished Lecturer. He has received many awards and recognitions, including Canada Research Chair, CRC Presidents Excellence Award, Canadian Federal Government Public Service Award, Ontario Early Researcher Award and five IEEE Best Paper Awards. He currently serves as an Editor/Associate Editor for IEEE Transactions on Communications, IEEE Transactions on Broadcasting, and IEEE Transactions on Vehicular Technology and he was also an Associate Editor for IEEE Transactions on Wireless Communications between 2007 and 2011, and IEEE Wireless Communications Letters between 2011 and 2016. Dr. Wang was involved in many IEEE conferences including GLOBECOM, ICC, VTC, PIMRC, WCNC, and CWIT, in different roles such as symposium chair, tutorial instructor, track chair, session chair and TPC co-chair.
\end{IEEEbiography}

\end{document}